\documentclass[12pt,draftclsnofoot,onecolumn]{IEEEtran}
\usepackage{graphicx}
\usepackage{amsmath}
\usepackage{amssymb}
\usepackage[caption=false]{subfig}
\usepackage[noadjust]{cite}
\usepackage{float}
\usepackage{color}
\usepackage{url}

\newtheorem{lemma}{Lemma}
\newtheorem{remark}{Remark}
\newtheorem{theorem}{Theorem}

\begin{document}

\bstctlcite{IEEEexample:BSTcontrol}

\title{NOMA for VLC Downlink Transmission with Random Receiver Orientation}

\author{Yavuz Yap{\i}c{\i} and \.{I}smail G\"{u}ven\c{c} 
\thanks{This work is supported in part by NSF CNS award 1422062.}
\thanks{The authors are with the Department of Electrical and Computer Engineering, North Carolina State University, Raleigh, NC (e-mail:~\{yyapici,iguvenc\}@ncsu.edu).}}

\maketitle


\begin{abstract}\vspace{-0.1in}
Visible light communications (VLC) is an emerging technology with a promise of viable solution to spectrum crunch problem in conventional radio frequency (RF) bands. In this work, we consider a downlink multiuser VLC network where users randomly change their location and vertical orientation. In order to increase the spectral efficiency, we consider the non-orthogonal multiple access (NOMA) transmission to serve multiple users simultaneously. In particular, we propose individual and group-based user ordering techniques for NOMA with various user feedback schemes. In order to reduce the computational complexity and link overhead, feedback on the channel quality is proposed to be computed using mean value of the vertical angle (instead of the exact instantaneous value), as well as the distance information. In addition, a two-bit feedback scheme is proposed for the group-based user scheduling, which relies on both the distance and vertical angle, and differs from the conventional one-bit feedback of the distance only. The outage probability and sum-rate expressions are derived analytically, which show a very good match with the simulation data. Numerical results verify that the practical feedback scheme with the mean vertical angle achieves a near-optimal sum-rate performance, and the two-bit feedback significantly outperforms the one-bit feedback.
\end{abstract}


\begin{IEEEkeywords}\vspace{-0.1in}
Non-orthogonal multiple access (NOMA), visible light communications (VLC), random receiver orientation, limited feedback, sum rates, outage probability.
\end{IEEEkeywords}


\section{Introduction}\label{sec:intro}

Visible light communications (VLC) is a promising technology for wireless 5G networks and beyond by leveraging the broad license-free optical spectrum at wavelengths of $380$-$750$ nm~\cite{Richardson13VLC}. Together with developments on light emitting diode (LED) as the primary illumination source, VLC networks appear as a viable solution for simultaneous illumination and communication at low power consumption and with high durability~\cite{Yapici2017MulEleVLC,Eroglu2018SofDef,Haas14VLCBeyond}. As recent research efforts reveal the power of VLC transmission being capable of achieving a speed of more than multiple Gigabits per second~\cite{Haas2014GbpsVLC}, this emerging technology enables ever increasing data-rate demanding mobile applications for next generation wireless networks. 

Towards improving the performance of multiuser VLC networks even more, a recent strategy of non-orthogonal multiple access (NOMA) appears as a powerful technology suggesting to serve multiple users at the same time and frequency slot, hence in a non-orthogonal fashion~\cite{Ding2017AppNoma, Poor2017NomaMul, Dobre2017PowDomNoma}. The NOMA strategy has been recently considered for VLC networks but with a limited attention. In~\cite{Uysal2015Noma}, NOMA is considered in a VLC scenario where the respective performance is compared to that of the orthogonal frequency division multiple access (OFDMA) scheme. The performance analysis of NOMA is conducted for VLC networks in \cite{Haas2016PerEvaNoma, Karagiannidis2016Noma} considering lighting quality and power allocation constraints. For a VLC NOMA system, a multiple-input multiple-output (MIMO) setting is explored in~\cite{Du2017OnPeMimo}, bit-error-rate (BER) analysis is performed in~\cite{Karagiannidis2017OnPerNoma}, sum-rate maximization is conducted in~\cite{Li2017FaiNoma}, a location based user grouping scheme is offered in~\cite{Xu2017UseGro}, and a phase pre-distorted  symbol detection method is proposed in~\cite{Cha2017JoiDet}.

VLC networks involving NOMA transmission have two main drawbacks which are i) overall complexity and overhead of the feedback mechanism involving channel quality, and ii) availability of the line-of-sight (LOS) link. Since the NOMA transmitter needs to order users according to their channel qualities, the channel gains (or related features) should be estimated at the user side, and then fed back to the transmitter. This mechanism comes with computational complexity while estimating unknown channel gains (or features), and results in link overhead while sending the channel quality information back to the transmitter. In addition, VLC transmission highly relies on LOS links, which may not be readily available all the time (e.g., when the receiving direction towards LED is outside the field-of-view (FOV) of the receiver).

The problem of LOS link being unavailable naturally arises in VLC networks with mobile users due to random receiver orientations, which is studied in~\cite{Haas2016AccPoiSel,Haas2017HanModInd,Yapici2017VerOriC,Yapici2017RanVerOr,Chen2017ImpBer,Wang2017ImpRec,Haas2018ImpTerOri,Haas2018ModRanOri}. In \cite{Haas2016AccPoiSel}, a new metric for the access point selection problem is proposed for receivers with random orientations. The handover problem for a VLC downlink network is considered in \cite{Haas2017HanModInd}, where the handover probability is computed by taking into account the device rotation. In \cite{Yapici2017VerOriC}, a general framework for random receiver orientation is developed where the square-channel gain distribution is derived analytically. The proposed framework applies to any prior distribution for the receiver orientation, and the approach is generalized to multi-LED scenarios in \cite{Yapici2017RanVerOr}. The impact of tilting the receiver angle on the BER performance is considered in \cite{Chen2017ImpBer}, and further studied in \cite{Wang2017ImpRec} to yield capacity bounds. Finally, the distribution of the receiver orientation in light-fidelity (LiFi) downlink networks is evaluated in \cite{Haas2018ImpTerOri,Haas2018ModRanOri} through indoor measurements. 

In this paper, we consider a multiuser VLC network where the mobile users with random vertical orientations are served by the NOMA strategy with various feedback schemes. To the best of our knowledge, this realistic VLC scenario with the NOMA transmission has not been studied in the literature before. Our specific contributions are summarized as follows.
\begin{itemize}
\item[i.] We propose a general framework for a multiuser VLC downlink scenario involving mobile users with varying horizontal distance and vertical receiver orientation. The distance and vertical angle are assumed to follow \textit{independent} random processes. Although the derivations and numerical results are presented for specific distance and angle distributions, the underlying methodology can be applied to a wide range of real-life scenarios (e.g., \cite{Haas2018ModRanOri}).
\item[ii.] We consider the NOMA transmission with the \textit{individual} and \textit{group-based} user scheduling techniques to serve multiple users simultaneously. While the transmitter picks up user pairs by considering each user \textit{one by one} in the individual user scheduling, user pairs are chosen from clusters in the group-based user scheduling. We propose a \textit{two-bit feedback} scheme to form clusters, which employs both the distance and vertical angle information, and differs from the conventional \textit{one-bit} feedback involving distance only (e.g., \cite{Ding2017RanBea}) as such.
\item[iii.] In order to relieve the computational complexity and link overhead with the feedback mechanisms involving \textit{instantaneous} vertical angle, we propose to use the \textit{mean} vertical angle instead as a limited yet sufficient feedback scheme. Moreover, we derive the analytical expressions for the outage probability and sum rates for each NOMA strategy under consideration, which show a very good match with the respective simulation data. 
\end{itemize}
 
The rest of the paper is organized as follows. Section~\ref{sec:system} introduces the system model. Section~\ref{sec:noma} considers various NOMA strategies in VLC networks, where the respective outage analysis is given in Section~\ref{sec:outage_analysis}. The numerical results are presented in Section~\ref{sec:results}, and the paper concludes with some final remarks in Section~\ref{sec:conclusion}.


\vspace{-0.15in}
\section{System Model} \label{sec:system} 
\begin{figure}[!t]
\centering
\includegraphics[width=0.55\textwidth]{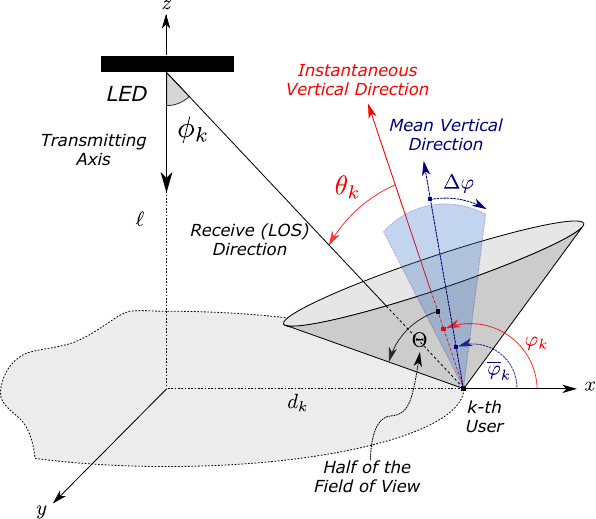}\vspace{-0.25in}
\caption{Multiuser VLC network showing $k$th user explicitly participating into NOMA transmission.}
\label{fig:setting}\vspace{-0.35in}
\end{figure}

We consider an indoor VLC downlink transmission scenario involving a single transmitting LED and $K$ users each of which is equipped with a photodetector for signal reception. The interaction between the LED and the $k$th user over a LOS link is depicted in Fig.~\ref{fig:setting}, and respective optical direct current (DC) channel gain is represented as~\cite{Haas2018PhySec}
\begin{align}\label{eqn:channel}
h_k =\frac{(m+1)A_r}{2\pi (\ell^2 + d_k^2)}\cos^m( \phi_k ) \cos( \theta_k ) \, \Pi\big[ \left| \theta_k \right| /\Theta \big],
\end{align}
where $\ell$ is the vertical distance between the LED and the plane including all the users, $d_k$ is the horizontal distance of the $k$th user to the LED, $\phi_k$ and $\theta_k$ are the angle of irradiance and incidence, respectively, of the $k$th user. The Lambertian order is denoted by $m\,{=}\,{-}1/\log_2(\cos(\Phi))$ with $\Phi$ being the half-power beamwidth of the LED, and $A_r$ and $\Theta$ are the detection area and {\color{black}half of the} FOV for the photodetectors, respectively. The function $\Pi[x]$ takes $1$ whenever $|x| \leq 1$, and is $0$ otherwise. Hence, \eqref{eqn:channel} is nonzero only if $\theta_k$ is smaller than $\Theta$ (i.e., LED is inside the FOV).   

We assume that the users are non-static within both the horizontal and vertical planes such that they are continuously changing their horizontal locations and vertical orientations. In particular, the horizontal distance $d_k$ of the $k$th user (representing the location in the horizontal plane) is assumed to follow a Uniform distribution with $\mathcal{U}\,[d_{\rm min}{,}\,d_{\rm max}]$ and $\Delta d \,{=}\,d_{\rm max}{-}d_{\rm min}$. In addition, the vertical orientation of the $k$th user is also varying around a \textit{mean vertical angle} $\overline{\varphi}_{k}$, which is picked up from a Uniform distribution with $\mathcal{U}\,[\overline{\varphi}_\text{min}{,}\,\overline{\varphi}_\text{max}]$ and $\Delta\overline{\varphi}\,{=}\,\overline{\varphi}_\text{max}{-}\overline{\varphi}_\text{min}$\footnote{\color{black}Although various distributions might be considered for the distance and/or angle (to comply with specifics of the communications scenario), this work prefers uniform distribution without any loss of generality since the fundamental findings herein are represented in terms of generic distribution functions, and, hence, can easily be extended to any particular distribution of interest.}. As a result, the $k$th user's orientation or, equivalently, the \textit{vertical angle} $\varphi_{k}$ takes a value from a uniform distribution with $\mathcal{U}\,[\overline{\varphi}_{k}{-}\Delta\varphi{,}\,\overline{\varphi}_{k}{+}\Delta\varphi]$ for a given value of $\overline{\varphi}_k$ and a maximum deviation angle of $\Delta\varphi$. Note that although $\varphi_{k}$ is represented in Fig.~\ref{fig:setting} as the the angle from $x$-axis, it actually describes an angle in the vertical domain, and hence is referred to as the vertical angle.

As an important distinction from the framework introduced in~\cite{Yapici2017RanVerOr}, we incorporate the effect of the distance $d_k$ on the incidence angle $\theta_k$ explicitly, which is given as
\begin{align}\label{eqn:incidenceang}
\theta_k = \pi - \tan^{{-}1}(\ell/d_k) - \varphi_k ,
\end{align}
{\color{black}where the horizontal distance $d_k$ and vertical angle $\varphi_k$ take values independently (in contrast to \cite{Yapici2017RanVerOr} where $d_k$ and $\varphi_k$ are coupled through the incidence angle)}. Note that $\theta_k$ in \eqref{eqn:incidenceang} is allowed to take either positive or negative values depending on {\color{black}the horizontal distance $d_k$ and} the vertical angle $\varphi_k$. This definition enables a more realistic scenario, where the vertical orientation can take any value regardless of how far the $k$th user is away from the LED, which is represented by the horizontal distance $d_k$ in the $xy$-plane. Hence, the incidence angle $\theta_k$ possesses the \textit{independent} contributions of the horizontal distance $d_k$ and the vertical angle $\varphi_k$.

Furthermore, we assume that the distance $d_k$, mean vertical angle $\overline{\varphi}_{k}$, and vertical angle $\varphi_{k}$ of the $k$th user do not change during a single transmission period over which the respective user rates are evaluated. In subsequent transmission periods, all these variables take new values from their respective distributions. We also assume that $d_k$ and $\overline{\varphi}_{k}$ are varying much slowly as compared to $\varphi_{k}$, and, hence, have relatively large coherence time. This assumption well aligns with realistic scenarios~{\color{black}\cite{Haas2018ModRanOri}}, where each user is changing its location and mean vertical direction slowly, {\color{black}whereas relatively small variations happen much quickly in actual vertical direction}. 

In this work, we consider the combination of $d_k$ and $\overline{\varphi}_{k}$ as a good candidate representing the channel quality, or equivalently the full channel state information (CSI), of the $k$th user. Since $d_k$ and $\overline{\varphi}_{k}$ can both be tracked with less computational burden {\color{black}(as compared to $\varphi_k$)}, we make use of these parameters while developing \textit{limited-feedback schemes} for the NOMA transmission. On the other hand, any limited-feedback scheme involving $d_k$ and $\overline{\varphi}_{k}$ (instead of $\varphi_k$) is likely to degrade the user rates as compared to the full CSI feedback. This is, in part, because the combination of $d_k$ and $\overline{\varphi}_{k}$ is not capable of capturing the true status of the receive direction (i.e., being inside or outside the FOV), which we call \textit{FOV status} and represented by $\Pi\left[ |\theta_k|/\Theta \right]$ in \eqref{eqn:channel}. In the subsequent sections, we consider this compromise between the limited-feedback (with lower computational burden and overhead) and the full CSI (with better user rates) mechanisms for the NOMA transmission.

\vspace{-0.15in}
\section{NOMA in VLC Downlink Channels} \label{sec:noma}
In this section, we consider the NOMA transmission for a VLC downlink scenario with full CSI and limited-feedback schemes. In particular, we first describe a general expression for sum rates, and then introduce two different NOMA strategies with various user scheduling techniques.

\vspace{-0.15in}
\subsection{Sum Rates for VLC NOMA Transmission}\label{sec:noma_userrates}
In NOMA transmission, multiple users having sufficiently \textit{distinct} channel qualities are served simultaneously using the same spectral (i.e., time-frequency) resources, which is hence classified as a \textit{non-orthogonal} transmission strategy. The respective messages of users paired for NOMA transmission, which are therefore referred to as NOMA users, are weighted by suitable power allocation coefficients. In general, each NOMA user is allocated with a power level which is inversely proportional to the channel quality of that user. The weighted messages are then combined together along with the superposition coding principle, and broadcast to all users having nonzero channel gain (i.e., with receive direction within the FOV). 

At the receiver side, each NOMA user decodes its own message after decoding, if any, messages of relatively weaker users (allocated with more power), while treating the messages of stronger users (allocated with less power) as noise. In the meanwhile, the decoded messages of the relatively weaker users are subtracted from the received signal, and the overall process is known as successive interference cancellation (SIC).

Without any loss of generality, we assume that the users are ordered in ascending order based on their indices such that $i$th user has the $i$th smallest nonzero channel gain. Moreover, we assume that $L$ users are involved in NOMA transmission out of $K$ users with $L\,{\leq}\,K$, and that $\mathcal{S}$ is the set including indices of these $L$ NOMA users. The signal-to-interference-plus-noise ratio (SINR) at the $j$th user while decoding the message of a weaker $i$th user with $i\,{<}\,j\,{\leq}\,K$ is 
\begin{align}\label{eqn:sinr_ij}
\mathsf{SINR}_{i{\rightarrow}j} = \frac{h_j^2 \beta_i^2}{h_j^2 \sum\limits_{k\in\mathcal{S}_i}\beta_k^2 + \gamma^{{-}1}},
\end{align}
where $\mathcal{S}_i$ is the set involving indices of the users being stronger than the $i$th user, $\beta_k$ is the optical power allocation coefficient of the $k$th user such that $\sum_{k{\in}\mathcal{S}}\beta_k^2\,{=}\,1$, and $\gamma$ is the equivalent electrical transmit signal-to-noise ratio (SNR). Note that \eqref{eqn:sinr_ij} implicitly assumes that the message of any $k$th user with $k\,{<}\,i$ (i.e., having a relatively weaker channel gain) has already been decoded successfully {\color{black}(i.e., perfect SIC)}, and subtracted from the received signal following SIC operation. Furthermore, the SINR of the $j$th user while decoding its own message is given as
\begin{align}\label{eqn:sinr_j}
\mathsf{SINR}_{j} = \frac{h_j^2 \beta_j^2}{h_j^2 \sum\limits_{k\in\mathcal{S}_j}\beta_k^2 + \gamma^{{-}1}},
\end{align}
where $\mathsf{SINR}_{\kappa}\,{=}\,h_\kappa^2 \beta_\kappa^2 \gamma$ for $\kappa$ being the index of the strongest NOMA user (i.e., $\mathcal{S}_\kappa$ is an empty set). At any NOMA user, the overall decoding mechanism is assumed to be in outage if instantaneous user rates associated with either of \eqref{eqn:sinr_ij} or \eqref{eqn:sinr_j} do not meet the respective target rates of the NOMA users based on their preset quality-of-service (QoS) requirements. 

Note that the conventional Shannon formulation does not hold for VLC links since the optical signal has certain average and peak power constraints as well as being non-negative. We therefore consider $R_{i{\rightarrow}j}\,{=}\frac{1}{2}\,\log_2 \left( 1 \,{+}\, \frac{e}{2\pi}\mathsf{SINR}_{i{\rightarrow}j} \right)$ and $R_{j}\,{=}\,\frac{1}{2}\log_2 \left( 1 \,{+}\, \frac{e}{2\pi}\mathsf{SINR}_{j} \right)$  as an instantaneous achievable rate pair associated with \eqref{eqn:sinr_ij} and \eqref{eqn:sinr_j}, respectively,~\cite{Moser2009OnCap, Haas2018PhySec}. The outage probability of the $j$th user is therefore given as 
\begin{align}\label{eqn:outage_j_1}
\mathsf{P}_{j}^{\rm o}  = 1 - \Pr \Big( \bigcap\limits_{k\in\overline{\mathcal{S}}_j} R_{k{\rightarrow}j} > \overline{R}_k, \, R_{j} > \overline{R}_{j}\Big),
\end{align}
where $\overline{R}_k$ is the QoS based target rate of the $k$th user, and $\overline{\mathcal{S}}_j$ is the set involving indices of the users being weaker than the $j$th user. Defining $\epsilon_k\,{=}\,\left( 2^{2{\overline{R}_k}}\,{-}\,1 \right) { \frac{2\pi}{e}}$, \eqref{eqn:outage_j_1} becomes
\begin{align}\label{eqn:outage_j_2}
\mathsf{P}_{j}^{\rm o}  = 1 - \Pr \Big( \bigcap\limits_{k\in\overline{\mathcal{S}}_j} \mathsf{SINR}_{k{\rightarrow}j} > \epsilon_k, \, \mathsf{SINR}_{j} > \epsilon_{j}\Big),
\end{align}
and, the respective sum-rate expression is accordingly defined as
\begin{align}\label{eqn:sumrate_noma}
R^{\rm NOMA} = \displaystyle\sum_{k{=}1}^{L} \left( 1 - \mathsf{P}_{k}^{\rm o} \right) \overline{R}_k .
\end{align}

For the orthogonal multiple access (OMA) transmission, all resources are allocated to a single user being served during $1/L$ of the transmission period, and hence the sum-rate expression is
\begin{align}\label{eqn:sumrate_oma}
R^{\rm OMA} = \displaystyle\sum_{k{=}1}^{L} \left[ 1 - \Pr \left( \left| h_k \right|^2 \leq \left( 2^{2{\overline{R}_k}}\,{-}\,1 \right) \frac{2\pi}{e} \right) \right] \overline{R}_k .
\end{align}

\vspace{-0.15in}
\subsection{Individual User Scheduling with Limited Feedback}\label{sec:noma_feedback_individual}
In this NOMA strategy, users to be served simultaneously are chosen among a total of $K$ users based on their \textit{individual} channel qualities. It is therefore vital for the NOMA transmitter to order \textit{all} the potential users according to their channel qualities based on the information transmitted back from the users. The optimal strategy from this perspective is therefore to order the users based on their full CSI given as follows
\begin{align}\label{eqn:order_fullcsi}
\left| h_1 \right|^2 \leq \left| h_2 \right|^2 \leq \dots \leq \left| h_K \right|^2.
\end{align}  

Because the channel gains need to be tracked continuously to enable the full CSI feedback, it is generally not practical to employ the order in \eqref{eqn:order_fullcsi}. Note that the \textit{instantaneous} value of the vertical angle $\varphi_k$ is changing relatively faster as compared to the distance $d_k$ and the mean vertical angle $\overline{\varphi}_{k}$, as discussed in Section~\ref{sec:system}. We therefore consider a limited-feedback mechanism, where each user sends its distance $d_k$ and the mean vertical angle $\overline{\varphi}_{k}$ information back to the NOMA transmitter. This mechanism circumvents the necessity of the continuous tracking of the vertical angle $\varphi_{k}$, and hence relieves the respective computational complexity at the user side. 

Based on this limited-feedback mechanism, the transmitter employs the following order while choosing the NOMA users
\begin{align}\label{eqn:order_limited}
\left| \overline{h}_{\vartheta_1} \right|^2 \leq \left| \overline{h}_{\vartheta_2} \right|^2 \leq \dots \leq \left| \overline{h}_{\vartheta_K} \right|^2,
\end{align} 
where $\overline{h}_{\vartheta_k}$ is the \textit{average} DC channel gain of the user with the index $\vartheta_k$, which corresponds to the $k$th smallest average DC channel gain among all. This definition of the average channel gain therefore implies that $\overline{h}_{\vartheta_k}$ might not necessarily appear as the $k$th record according to the full CSI feedback based order of \eqref{eqn:order_fullcsi}. Substituting the vertical angle $\varphi_k$ with its mean $\overline{\varphi}_k$ in the original channel expression of \eqref{eqn:channel}, and using the definition of the incidence angle given in \eqref{eqn:incidenceang}, the average DC channel gain for user $k$ with $k\,{\in}\,\left\lbrace \vartheta_1,\vartheta_2,\dots,\vartheta_K\right\rbrace$ is given as follows
\begin{align}\label{eqn:channel_mean}
\overline{h}_k =\frac{(m+1)A_r}{2\pi (\ell^2 + d_k^2)}\cos^m( \phi_k ) \left| \cos(\tan^{{-}1}(\ell/d_k) + \overline{\varphi}_k ) \right| \, \Pi\big[ \left| \pi - \tan^{{-}1}(\ell/d_k) - \overline{\varphi}_k \big| /\Theta \right],
\end{align}
where the absolute value of $\cos\left( \tan^{{-}1}(\ell/d_k) + \overline{\varphi}_k \right)$ guarantees the non-negativity of the gain. 

\begin{remark}
The FOV status, which represents the status of the receive direction (from the LED to the user) being inside the FOV or not, is a function of the distance $d_k$ and the vertical angle $\varphi_k$, as captured by $\Pi$ function in \eqref{eqn:channel}. As a result, any limited-feedback mechanism solely relying either on the distance $d_k$ or the mean vertical angle $\overline{\varphi}_k$ (as a substitute for $\varphi_k$) cannot capture the FOV status correctly. Hence, the respective sum-rate performance might degrade since the FOV status is one of the main contributors to the channel gain. On the other hand, although the proposed feedback mechanism employing $d_k$ and $\overline{\varphi}_k$ together is still not capable of having the correct FOV status precisely, the discrepancy becomes marginal as the maximum deviation angle $\Delta\varphi$ (between the mean angle $\overline{\varphi}_k$ and the maximum instantaneous value $\varphi_k$) gets smaller. 
\end{remark}

\vspace{-0.15in}
\subsection{Group-Based User Scheduling with Two-Bit Feedback}\label{sec:noma_feedback_group}
We now consider a different NOMA strategy where the transmitter does not order all $K$ potential users individually (as in Section~\ref{sec:noma_feedback_individual}), but rather groups them based on \textit{two-bit} information on their channel qualities. This strategy therefore employs a low-rate feedback mechanism, where the user $k$ computes its feedback bits based on one of the following channel features: 1) the distance $d_k$ and the instantaneous vertical angle $\varphi_{k}$, or 2) the distance $d_k$ and the mean vertical angle $\overline{\varphi}_{k}$. In either case, the distance and (average) incidence angle are compared to their own preset threshold values, and the result is transmitted back to the transmitter in two bits of information. Note that the feedback bit based on $\overline{\varphi}_{k}$ circumvents the continuous tracking of $\varphi_{k}$, and therefore suggests a more practical feedback scheme, as discussed in Section~\ref{sec:noma_feedback_individual}.

More specifically, the respective feedback bits are $\left\lbrace \Pi \left( d_k/d_{\rm th} \right),\Pi \left( \left|\theta_k\right|/\theta_{\rm th} \right) \right\rbrace$ if $d_k$ and $\varphi_{k}$ are both available to the user $k$ for feedback computation. In this formulation, $\theta_k$ is the incidence angle given by \eqref{eqn:incidenceang}, $d_{\rm th}$ and $\theta_{\rm th}$ are threshold values, and $\Pi$ function is defined along with \eqref{eqn:channel}. In addition, if the user $k$ has the information of $d_k$ and $\overline{\varphi}_{k}$, then the feedback bits become $\left\lbrace \Pi \left( d_k/d_{\rm th} \right),\Pi \left( \left|\overline{\theta}_k\right|/\theta_{\rm th} \right) \right\rbrace$ with $\overline{\theta}_k \,{=}\, \pi {-} \tan^{{-}1}(\ell/d_k) {-} \overline{\varphi}_k$.

At the transmitter side, potential users can only be distinguished from the others based on this two-bit of feedback information. All users are therefore split into groups, where each group is composed of users having the same feedback bits. Assuming that the distance and instantaneous vertical angle are used in feedback computations, the groups of users having weaker and stronger channel gains can be represented, respectively, as follows
\begin{align}
\mathcal{S}_{\rm W, \varphi} &= \left\lbrace k \mid  d_k\,{>}\,d_{\rm th},\,    \left|\theta_k\right| \,{>}\, \theta_{\rm th} \right\rbrace, \label{eqn:set_weak_ins} \\
\mathcal{S}_{\rm S, \varphi} &= \left\lbrace k \mid  d_k\,{\leq}\,d_{\rm th},\, \left|\theta_k\right| \,{\leq}\,\theta_{\rm th} \right\rbrace, \label{eqn:set_strong_ins}
\end{align}
which make use of the channel gain expression in \eqref{eqn:channel}. Similarly, whenever the distance and mean vertical angle are used while computing feedback bits, the groups of users having weaker and stronger channel gains are given, respectively, as follows  
\begin{align}
\mathcal{S}_{\rm W, \overline{\varphi}} &= \left\lbrace k \mid  d_k\,{>}\,d_{\rm th},\, \left|\overline{\theta}_k \right| \,{>}\,\theta_{\rm th} \right\rbrace,  \label{eqn:set_weak_mean} \\
\mathcal{S}_{\rm S, \overline{\varphi}} &= \left\lbrace k \mid  d_k\,{\leq}\,d_{\rm th},\, \left|\overline{\theta}_k \right|  \,{\leq} \,\theta_{\rm th} \right\rbrace, \label{eqn:set_strong_mean}
\end{align}
which consider the \textit{average} channel gain expression in \eqref{eqn:channel_mean}.

Note that the goal of the NOMA transmitter is to pair relatively weaker and stronger users in order to provide sufficient user separation in the power domain for a successful SIC decoding. {\color{black}We, therefore, confine our search of NOMA users to the sets $\mathcal{S}_{\rm W, \varphi}$ ($\mathcal{S}_{\rm W, \overline{\varphi}}$) and $\mathcal{S}_{\rm S, \varphi}$ ($\mathcal{S}_{\rm S, \overline{\varphi}}$), which are more likely to involve good candidates for weak and strong NOMA users, respectively.} Although users are not ordered individually based on their channel qualities, there is actually an order between channel gains of the users in $\mathcal{S}_{\rm W, \varphi}$ ($\mathcal{S}_{\rm W, \overline{\varphi}}$) and $\mathcal{S}_{\rm S, \varphi}$ ($\mathcal{S}_{\rm S, \overline{\varphi}}$). Based on the available two-bit feedback information, we therefore pick up weaker NOMA user either from $\mathcal{S}_{\rm W, \varphi}$ or $\mathcal{S}_{\rm W, \overline{\varphi}}$, and stronger NOMA user is similarly chosen either from $\mathcal{S}_{\rm S, \varphi}$ or $\mathcal{S}_{\rm S, \overline{\varphi}}$. 

\begin{remark}
Although we consider two users involved in this NOMA strategy, more users can still be picked up within this framework at the expense of loosing practicality and degraded power domain separation (i.e., worse SIC decoding performance). Furthermore, another low-rate feedback scheme can also be considered within this framework, where one-bit feedback conveys only the distance information (i.e., whether $d_k$ is greater than a threshold $d_{\rm th}$). This feedback scheme, which is initially considered in~\cite{Poor2017RanBea} for millimeter-wave channels, does not capture the FOV status at all, and the resulting sum-rate performance is hence much worse than that of the two-bit feedback schemes considered herein, as shown through numerical results of Section~\ref{sec:results}. 
\end{remark}

\vspace{-0.15in}
\section{Outage Analysis for VLC NOMA}\label{sec:outage_analysis}
In this section, we derive the exact outage probability expressions for the NOMA strategies in Section~\ref{sec:noma}. In the analysis, we assume that the NOMA transmission serves two users at a time, although the results can be generalized to scenarios with more than two NOMA users.

\vspace{-0.15in}
\subsection{Outage Formulation}\label{sec:outage_formulation}
In the VLC downlink transmission, the channel gain in \eqref{eqn:channel} can take either zero or a nonzero value, which depends on the receive direction being inside or outside the FOV. This is a major difference of the VLC transmission from its RF counterparts, and hence we force the NOMA transmitter to schedule only the users having nonzero channel gains. Following the convention of Section~\ref{sec:noma_userrates}, we designate $i$ and $j$ being the index of the users having weaker and stronger \textit{nonzero} channel gains, respectively, with $i\,{<}\,j\,{\leq}\,K$. Note that the NOMA transmission described in Section~\ref{sec:noma_feedback_individual} starts only if we have at least $j$ users having nonzero channel gain. On the other hand, the strategy of Section~\ref{sec:noma_feedback_group} requires that there should be at least one user with nonzero channel gain in both the sets $\mathcal{S}_{\rm W, \varphi}$ ($\mathcal{S}_{\rm W, \overline{\varphi}}$) and $\mathcal{S}_{\rm S, \varphi}$ ($\mathcal{S}_{\rm S, \overline{\varphi}}$) to start the NOMA transmission.

Considering the NOMA strategy in Section~\ref{sec:noma_feedback_individual}, the outage probability for the $i$th user is 
\begin{align}
\mathsf{P}_{i}^{\rm o}  &= 1 - \Pr \left( \mathsf{SINR}_{i} > \epsilon_i \mid K_{\rm nz} \geq j \right)  \label{eqn:outpr_i_1} \\
&= 1 - \Pr \left( \frac{h_i^2 \beta_i^2}{h_i^2 \beta_j^2 + \gamma^{{-}1}} > \epsilon_i \mid K_{\rm nz} \geq j \right), \label{eqn:outpr_i_2}\\
&= 1 - \Pr \left( h_i^2 > \eta_i \mid K_{\rm nz} \geq j \right),  \label{eqn:outpr_i_3}
\end{align}
where $\eta_i\,{=}\,\frac{\epsilon_i/\gamma}{\beta_i^2{-}\beta_j^2\epsilon_i}$, and $K_{\rm nz}$ is the discrete random variable representing the number of users having nonzero channel gain. Similarly, the outage probability for the $j$th user is given as
\begin{align}
\mathsf{P}_{j}^{\rm o}  &= 1 - \Pr \left( \mathsf{SINR}_{i{\rightarrow}j} > \epsilon_i, \, \mathsf{SINR}_{j} > \epsilon_{j} \mid K_{\rm nz} \geq j\right), \label{eqn:outpr_j_1}\\
&= 1 - \Pr \left( \frac{h_j^2 \beta_i^2}{h_j^2 \beta_j^2 + \gamma^{{-}1}} > \epsilon_i, \, h_j^2 \beta_j^2 \gamma > \epsilon_j \mid K_{\rm nz} \geq j\right), \label{eqn:outpr_j_2}\\
&= 1 - \Pr \left( h_j^2 > \eta_j \mid K_{\rm nz} \geq j\right),  \label{eqn:outpr_j_3}
\end{align}
where $\eta_j\,{=}\,\max\left\lbrace \frac{\epsilon_i/\gamma}{\beta_i^2{-}\beta_j^2\epsilon_i} ,\, \frac{\epsilon_j/\gamma}{\beta_j^2} \right\rbrace$. Finally, employing \eqref{eqn:outpr_i_3} and \eqref{eqn:outpr_j_3} in \eqref{eqn:sumrate_noma} gives the outage sum rates for the NOMA transmission under consideration. 

For the NOMA strategy in Section~\ref{sec:noma_feedback_group}, we have the same expressions in \eqref{eqn:outpr_i_3} and \eqref{eqn:outpr_j_3}, except that we do not need to use the given condition $K_{\rm nz} \,{\geq}\, j$. Note that the respective requirement to start NOMA transmission is to have at least one user in each of the sets $\mathcal{S}_{\rm W, \varphi}$ ($\mathcal{S}_{\rm W, \overline{\varphi}}$) and $\mathcal{S}_{\rm S, \varphi}$ ($\mathcal{S}_{\rm S, \overline{\varphi}}$), which will individually be taken into account during derivations presented in Section~\ref{sec:cdf_twobit}. Note also that the outage probability expressions in \eqref{eqn:outpr_i_3} and \eqref{eqn:outpr_j_3} actually involves the conditional cumulative distribution function (CDF) of the nonzero square-channels, which will therefore be derived in the following for the NOMA strategies in Section~\ref{sec:noma_feedback_individual} and Section~\ref{sec:noma_feedback_group} separately. 

\vspace{-0.15in}
\subsection{Angle and User Distributions}\label{sec:angle_user_distribution}
In this section, we consider the distributions of the vertical angle $\varphi$ and the number of users having nonzero channel gain, denoted by $K_{\rm nz}$, which will be needed in the desired CDF derivations of the nonzero square-channel gain.

\begin{lemma}\label{lem:cdf_varphi}
Assuming that the unordered vertical angle $\varphi$ is a conditionally uniform random variable with $\mathcal{U}\,[\overline{\varphi}{-}\Delta\varphi{,}\,\overline{\varphi}{+}\Delta\varphi]$ for a given $\overline{\varphi}$, where the mean vertical angle $\overline{\varphi}$ follows a uniform distribution with $\mathcal{U}\,[\overline{\varphi}_\text{min}{,}\,\overline{\varphi}_\text{max}]$, the unconditional CDF of $\varphi$ is given as
\begin{align}\label{eqn:cdf_varphi}
F_{\varphi}(x) = \begin{cases}
1 & \textrm{ if } x \geq \overline{\varphi}_{\rm max} + \Delta\varphi,\\
\displaystyle \frac{\left(x+\Delta\varphi - \overline{\varphi}_{\rm min}\right)^2}{4 \Delta\varphi \Delta\overline{\varphi}} 
& \textrm{ if } \overline{\varphi}_{\rm min} - \Delta\varphi \leq x \leq \zeta_{\min}, \\
\displaystyle 1-\frac{\left( \overline{\varphi}_{\rm max} - x + \Delta\varphi \right)^2}{4 \Delta\varphi \Delta\overline{\varphi}}
 & \textrm{ if } \zeta_{\max} \leq x \leq \overline{\varphi}_{\rm max} + \Delta\varphi , \\
\displaystyle \frac{x-\overline{\varphi}_{\rm min}}{\Delta\overline{\varphi}} & \textrm{ if } \overline{\varphi}_{\rm min} + \Delta\varphi \leq x \leq \overline{\varphi}_{\rm max} - \Delta\varphi, \\
\displaystyle \frac{1}{2\Delta\varphi}\left( x+\Delta\varphi-\frac{\overline{\varphi}_{\rm min}+\overline{\varphi}_{\rm max}}{2} \right) & \textrm{ if } \overline{\varphi}_{\rm max} - \Delta\varphi \leq x \leq \overline{\varphi}_{\rm min} + \Delta\varphi, \\
0 & \textrm{ if } x < \overline{\varphi}_{\rm min} - \Delta\varphi,
\end{cases} 
\end{align}
where $\zeta_{\min} \,{=}\, \min \left(  \overline{\varphi}_{\rm min} + \Delta\varphi, \overline{\varphi}_{\rm max} - \Delta\varphi \right)$ and $\zeta_{\max} \,{=}\, \max \left(  \overline{\varphi}_{\rm min} + \Delta\varphi, \overline{\varphi}_{\rm max} - \Delta\varphi \right)$.
\end{lemma}

\begin{IEEEproof}
See Appendix~\ref{app:cdf_varphi_derivation}.
\end{IEEEproof}

\begin{lemma}\label{lem:numof_nzusers}
The random variable $K_{\rm nz}$, which denotes the number of users having nonzero channel gain, follows Binomial distribution with $\mathcal{B}(K,p)$. The success probability $p$ is given as 
\begin{align}
p &= \frac{1}{\Delta d} \int_{d_{\rm min}}^{d_{\rm max}} \Delta F_{\varphi} \left(r,\Theta \right) {\rm d}r,\label{eqn:success_prob}
\end{align}
where $\Delta F_{\varphi} \left( x,y \right) \,{=}\, F_{\varphi}\left( \pi {-} \tan^{{-}1}(\ell/x) {+} y \right) \,{-}\, F_{\varphi} \left( \pi {-} \tan^{{-}1}(\ell/x) {-} y  \right)$ with $F_{\varphi}$ given by \eqref{eqn:cdf_varphi}.
\end{lemma}
\begin{IEEEproof}
See Appendix~\ref{app:numof_nzusers}.
\end{IEEEproof}

Since $K_{\rm nz}$ is Binomial with $\mathcal{B}(K,p)$, the respective probability mass function (PMF) is
\begin{align}\label{eqn:pmf_knz}
p_{K_{\rm nz}}(k) = \Pr \left( K_{\rm nz} = k \right) = \binom{K}{k} p^k (1-p)^{K-k},
\end{align}
for $k \,{=}\, 1,2,\dots,K$. Note that since the NOMA transmission starts whenever there are at least $j$ users with nonzero channel gain, we apply the condition $K\,{\geq}\,j$ to the PMF in \eqref{eqn:pmf_knz} which yields
\begin{align}\label{eqn:pmf_knz_mdf}
p_{K_{\rm nz}} \left( k | k_{\rm min} \right) = \begin{cases}
\displaystyle c_{\rm nz} \binom{K}{k} p^k (1-p)^{K-k} & \textrm{ if } k \geq k_{\rm min},\\
0 & \textrm{ otherwise},
\end{cases}
\end{align}
where $c_{\rm nz} \,{=}\, \sum_{k=k_{\rm min}}^{K} \binom{K}{k} p^k (1-p)^{K-k} $, and $k_{\rm min}$ is the minimum number of users having nonzero channel gain to start the NOMA transmission (i.e., $k_{\rm min}\,{=}\,j$ in the strategy of Section~\ref{sec:noma_feedback_individual}).

\begin{figure}[!t]
\centering
\subfloat[The CDF of $\varphi$]{\includegraphics[width=0.5\textwidth]{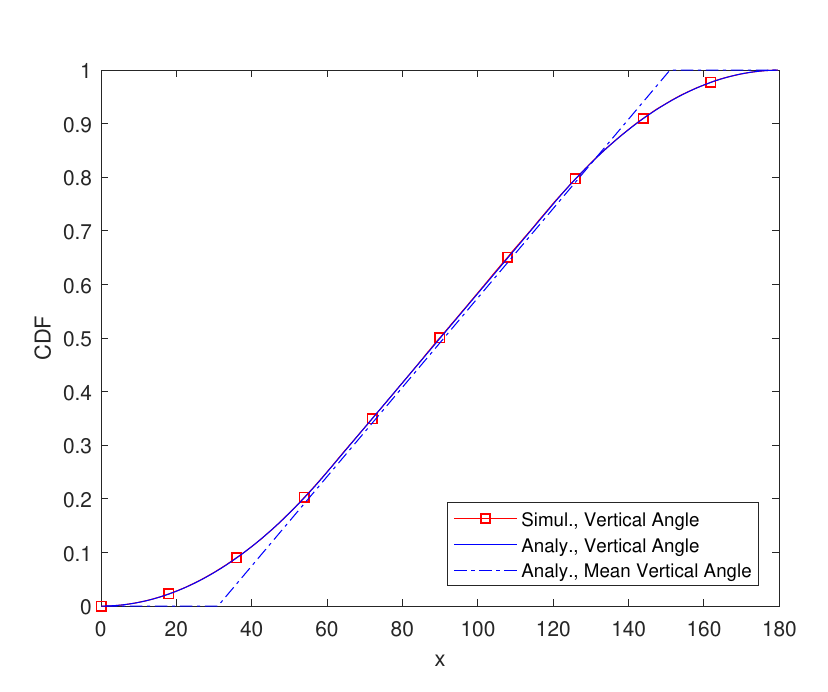}
\label{fig:cdf_vertang}}
\subfloat[The PMF of $K_{\rm nz}$]{\includegraphics[width=0.5\textwidth]{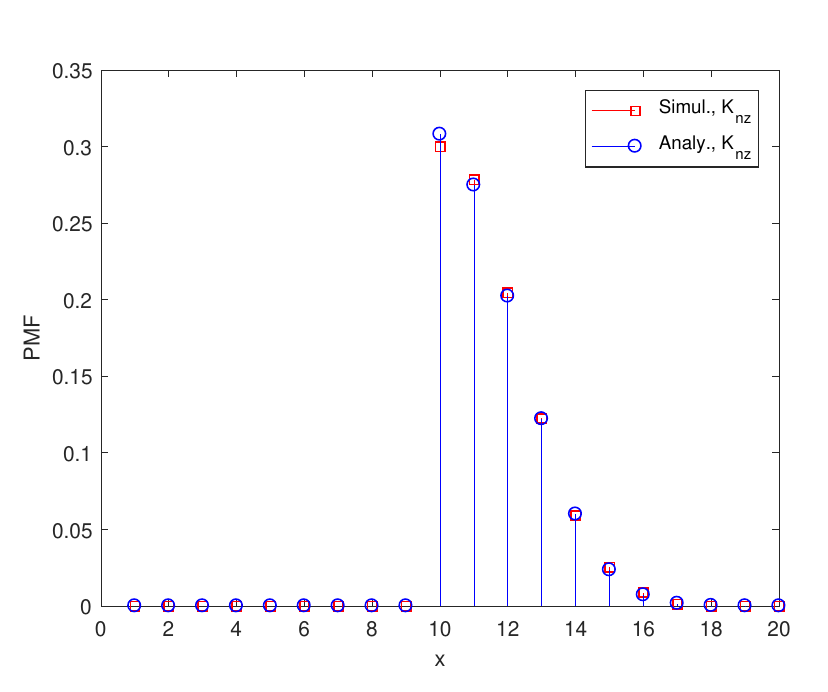}
\label{fig:pmf_knz}}
\caption{Analytical and simulation results for the CDF of the vertical angle $\varphi$ and the PMF of $K_{\rm nz}$ (the number of users having nonzero channel gain) for $\overline{\varphi}_\text{min}\,{=}\,30^\circ$, $\overline{\varphi}_\text{max}\,{=}\,150^\circ$, $\Delta\varphi\,{=}\,30^\circ$, $j\,{=}\,10$, $\Theta\,{=}\,{\color{black}60}^\circ$, and $K\,{=}\,20$.}
\label{fig:cdfvertang_pmfknz}\vspace{-0.3in}
\end{figure}

In Fig.~\ref{fig:cdfvertang_pmfknz}, we present Monte Carlo based simulation results verifying Lemma~\ref{lem:cdf_varphi} and Lemma~\ref{lem:numof_nzusers}. In particular, we assume $\overline{\varphi}_\text{min}\,{=}\,30^\circ$, $\overline{\varphi}_\text{max}\,{=}\,150^\circ$, $\Delta\varphi\,{=}\,30^\circ$, $j\,{=}\,10$, $\Theta\,{=}\,{\color{black}60}^\circ$, and $K\,{=}\,20$ for the setting of Section~\ref{sec:system}. We observe that the analytical expressions for the CDF of $\varphi$ and the PMF of $K_{\rm nz}$ (for $K\,{\geq}\,j$), given by \eqref{eqn:cdf_varphi} and \eqref{eqn:pmf_knz_mdf}, respectively, both follow the corresponding simulation data successfully. In addition, we also provide the CDF of the mean vertical angle in Fig.~\ref{fig:cdfvertang_pmfknz}\subref{fig:cdf_vertang} to highlight the effect of random variation in the instantaneous vertical angle $\varphi$ on top of the mean $\overline{\varphi}$. Note that, as the maximum deviation angle $\Delta\varphi$ goes to zero, the instantaneous value $\varphi$ will surely converge to the mean value $\overline{\varphi}$. 

\vspace{-0.15in}
\subsection{Square-Channel Distribution for Individual User Scheduling}\label{sec:cdf_individual}
In this section, we derive the unordered and ordered distributions of the nonzero square-channel gain assuming that all $K$ users appear individually in the order, as in Section~\ref{sec:noma_feedback_individual}. Furthermore, we assume that the full CSI is available to the transmitter for each user.

\begin{theorem}\label{the:cdf_unordered}
The CDF of the unordered nonzero square-channel is given as
\begin{align} 
F_{h^2|h>0}(x) &= 1 {-} \frac{\displaystyle \int_{d_{\rm min}}^{d_{\rm max}} \Delta F_{\varphi} \left( r,\psi \left(x,r,\Theta\right) \right) {\rm d}r}
{\displaystyle \int_{d_{\rm min}}^{d_{\rm max}} \Delta F_{\varphi} \left( r,\Theta \right) {\rm d}r}, \label{eqn:cdf_unordered}
\end{align}
where $\psi(x,y,z) \,{=}\, \min \left( 1/2\cos^{{-}1}\!\left( 2\min \left(x\upsilon(y),1\right){-}1\right),z \right)$ with $\upsilon(x)\,{=}\,(\ell^2 + x^2)^{m+2}h_c^{{-}2}$ and $h_c^2\,{=}\,(m+1)A_r\ell^m/2\pi$, and $\Delta F_{\varphi} \left( x,y \right)$ is defined in Lemma~\ref{lem:numof_nzusers}.
\end{theorem}
\begin{IEEEproof}
See Appendix~\ref{app:cdf_fullcsi}.
\end{IEEEproof}

Note that whenever we choose the $k$th user having nonzero channel gain, we actually order a total of $K_{nz}$ users based on their full CSI, since the remaining $K\,{-}\,K_{nz}$ users all have zero channel gain. Assuming a particular value $K_{nz}\,{=}\,n$, the conditional CDF of the ordered nonzero square-channel gain of the $k$th user can be found using order statistics~\cite{Nagaraja2005OrdSta}, which is given as
\begin{align}\label{eqn:cdf_ordered_givenknz}
F_{h_k^2|h_k>0}(x | K_{nz}\,{=}\,n) &= \sum\limits_{\ell=k}^{n} \binom{n}{\ell} \left[ F_{h^2|h>0}(x)\right]^{\ell} \left[ 1-F_{h^2|h>0}(x)\right]^{n-\ell},
\end{align}
where $F_{h^2|h>0}(x)$ is the unordered CDF in \eqref{eqn:cdf_unordered}. In order to obtain the respective unconditional CDF, we need to average \eqref{eqn:cdf_ordered_givenknz} over the distribution of $K_{nz}$ given in Lemma~\ref{lem:numof_nzusers}, which yields
\begin{align}
F_{h_k^2|h_k>0}(x) &= \sum\limits_{n=k_{\min}}^{K} p_{K_{\rm nz}} \left( n | k_{\rm min} \right) \sum\limits_{\ell=k}^{n} \binom{n}{\ell} \left[ F_{h^2|h>0}(x)\right]^{\ell} \left[ 1-F_{h^2|h>0}(x)\right]^{n-\ell},\\
&= \sum\limits_{n=k_{\min}}^{K} \sum\limits_{\ell=k}^{n} \binom{n}{\ell} \binom{K}{n} c_{\rm nz} \, p^{n} (1-p)^{K-n} \left[ F_{h^2|h>0}(x)\right]^{\ell} \left[ 1-F_{h^2|h>0}(x)\right]^{n-\ell},\label{eqn:cdf_ordered}
\end{align}
where $p_{K_{\rm nz}}$, $p$, and $c_{\rm nz}$ are all defined in Lemma~\ref{lem:numof_nzusers}. Finally, the desired outage probabilities for both weaker and stronger users, which are given by \eqref{eqn:outpr_i_3} and \eqref{eqn:outpr_j_3} in Section~\ref{sec:outage_formulation}, respectively, can be computed using \eqref{eqn:cdf_ordered} for the NOMA strategy presented in Section~\ref{sec:noma_feedback_individual}.

\begin{figure}[!t]
\centering
\subfloat[Individual User Scheduling, $j{=}10$]{\includegraphics[width=0.5\textwidth]{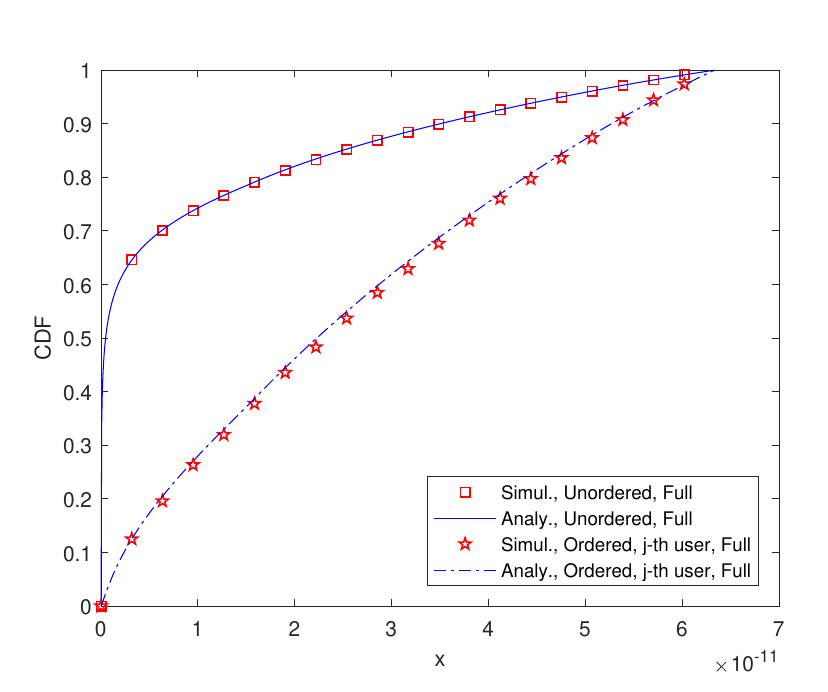}
\label{fig:cdf_squarechannel_ind}}
\subfloat[Group-Based User Scheduling, $d_{\rm th}{=}1$m, $\theta_{\rm th}{=}6^\circ$]{\includegraphics[width=0.5\textwidth]{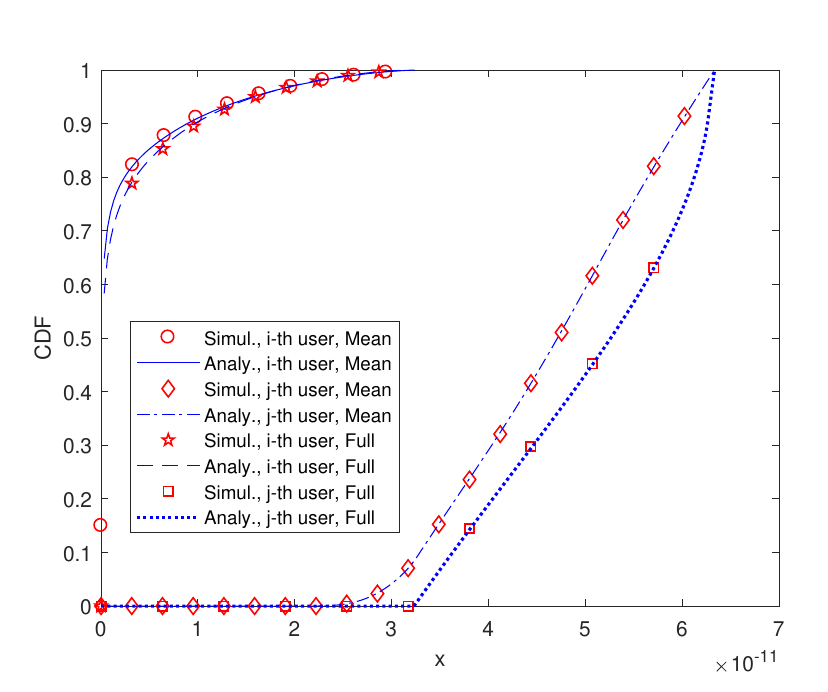}
\label{fig:cdf_squarechannel_two}}
\caption{Analytical and simulation results for the CDF of the nonzero square-channel for $\overline{\varphi}_\text{min}\,{=}\,30^\circ$, $\overline{\varphi}_\text{max}\,{=}\,150^\circ$, $\Delta\varphi\,{=}\,30^\circ$, $\Theta\,{=}\,{\color{black}60}^\circ$, and $K\,{=}\,20$ along with full CSI and mean vertical angle based limited-feedback schemes.}
\label{fig:cdf_squarechannel}\vspace{-0.3in}
\end{figure}

In Fig.~\ref{fig:cdf_squarechannel}\subref{fig:cdf_squarechannel_ind}, we provide numerical results verifying the unordered and ordered CDF derivation for the nonzero square-channel keeping the setting of Fig.~\ref{fig:cdfvertang_pmfknz} the same. We observe a very good match between analytical and simulation results, and that $j$th (stronger) user channel takes larger values as compared to arbitrary unordered user channel, as expected. 



As a final remark, we do not consider to derive the CDF of the nonzero square-channel gain involving the feedback of the \textit{mean} vertical angle, which is far too complex and requires lengthy derivations. In particular, we need to resort to the theory of \textit{concomitants} from the order statistics literature \cite{Nagaraja2009DisCon} in order to obtain a suitable closed form expression. We therefore leave the respective analysis to be subject of a future study. Note also that the sum-rate performance of the full CSI feedback is actually a very good representative for that of the mean vertical angle based feedback since they achieve a similar sum-rate performance, as presented in Section~\ref{sec:results}.

\vspace{-0.15in}
\subsection{Square-Channel Distribution for Group-Based User Scheduling}\label{sec:cdf_twobit}
In this section, we derive the distribution of the nonzero square-channel gain for the NOMA strategy based on two-bit feedback introduced in Section~\ref{sec:noma_feedback_individual}. In the following, we take into account two different feedback categories, where either the instantaneous vertical angle $\varphi$ or the mean vertical angle $\overline{\varphi}$ is available for feedback computation with the distance $d$ in either case.

Assuming that the two-bit feedback is computed using the distance $d$ and the instantaneous vertical angle $\varphi$, the weaker user $i$ and the stronger user $j$ are randomly chosen from the groups $\mathcal{S}_{\rm W, \varphi}$ and $\mathcal{S}_{\rm S, \varphi}$, respectively, which are given by \eqref{eqn:set_weak_ins} and \eqref{eqn:set_strong_ins}, respectively. The distribution of the nonzero square-channel gains are given accordingly in the following theorem.

\begin{theorem}\label{the:cdf_twobit_ins}
The CDF of the nonzero square-channel gain for user $i\,{\in}\,\mathcal{S}_{\rm W, \varphi}$ is given as
\begin{align} 
F_{h_i^2|h_i>0}(x) &= \frac{\displaystyle \int_{d^*(x)}^{d_{\rm max}} \Big( \Delta F_{\varphi} \left( r,\Theta \right) - \Delta F_{\varphi} \left( r,\omega \left(x,r,\theta_{\rm th}\right) \right) \Big) {\rm d}r}
{\displaystyle \int_{d_{\rm th}}^{d_{\rm max}} \Big( \Delta F_{\varphi} \left( r,\Theta \right) - \Delta F_{\varphi} \left( r,\theta_{\rm th} \right) \Big) {\rm d}r}, \label{eqn:cdf_twobit_ins_i}
\end{align}
where $\omega(x,y,z) \,{=}\, \max \left( 1/2\cos^{{-}1}\!\left( 2\min \left(x\upsilon(y),1\right){-}1\right),z \right)$ with $\upsilon(x)$ defined in Theorem~\ref{the:cdf_unordered}, $d^*(x) \,{=}\, u \left( \sqrt[]{\left( h_c^2 \cos^2\Theta/x \right)^{1/(m{+}2)}{-}\ell^2}, d_{\rm th}, d_{\max} \right)$ with $u(x,y,z) \,{=}\, \min \left( \max \,( x, y ), z \right)$, and $\Delta F_{\varphi} \left( x,y \right)$ is given in Lemma~\ref{lem:numof_nzusers}.

Similarly, the CDF of the nonzero square-channel gain for user $j\,{\in}\,\mathcal{S}_{\rm S, \varphi}$ is given as
\begin{align} \label{eqn:cdf_twobit_ins_j}
F_{h_j^2|h_j>0}(x) &= 1 {-} \frac{\displaystyle \int_{d_{\rm min}}^{d_{\rm th}} \Delta F_{\varphi} \left( r,\psi \left(x,r,\theta_{\rm th}\right) \right) {\rm d}r}
{\displaystyle \int_{d_{\rm min}}^{d_{\rm th}} \Delta F_{\varphi} \left( r,\theta_{\rm th} \right) {\rm d}r}, 
\end{align}
where $\psi(x,y,z)$ is given in Theorem~\ref{the:cdf_unordered}{\color{black}, and $\Delta F_{\varphi} \left( x,y \right)$ is given in Lemma~\ref{lem:numof_nzusers}}. 
\end{theorem}
\begin{IEEEproof}
See Appendix~\ref{app:cdf_twobit_ins}.
\end{IEEEproof}

When the two-bit feedback is computed using the distance $d$ and the mean vertical angle $\overline{\varphi}$, the weaker and stronger users are chosen randomly from the sets given by \eqref{eqn:set_weak_mean} and \eqref{eqn:set_strong_mean}, respectively, i.e., $i \,{\in}\, \mathcal{S}_{\rm W, \overline{\varphi}}$ and $j \,{\in}\, \mathcal{S}_{\rm S, \overline{\varphi}}$. The respective distributions are given in the next theorem.

\begin{theorem}\label{the:cdf_twobit_mean}
The CDF of the nonzero square-channel for user $i\,{\in}\,\mathcal{S}_{\rm W, \overline{\varphi}}$ is given as
\begin{align} \label{eqn:cdf_twobit_mean_i}
F_{h_i^2|\overline{h}_i>0}(x) = \frac{\displaystyle A(d^*(x))}{A(d_{\rm th})} + \frac{1}{\Delta\overline{\varphi} A(d_{\rm th})} \int_{d_{\rm th}}^{d^*(x)} \int_{\mathcal{S}_{\overline{\varphi}}(r)} \Big( 1 - \Delta F_{\varphi|\overline{\varphi}} \left( r,\Psi \left(x,r,\Theta\right) \right) \Big) {\rm d}\overline{\varphi} \, {\rm d}r ,
\end{align}
where $\Psi(x,y,z) \,{=}\, \min \left( 1/2\cos^{{-}1}\!\left( 2x\upsilon(y){-}1\right),z \right)$, $d^*(x) \,{=}\, u \Big( \sqrt[]{\left( h_c^2/x \right)^{1/(m{+}2)}{-}\ell^2} , d_{\rm th}, d_{\max}\Big)$ with $u(x,y,z)$ given in Theorem~\ref{the:cdf_twobit_ins}, and $\Delta F_{\varphi|\overline{\varphi}} \left( x,y \right) {=}\, F_{\varphi|\overline{\varphi}}\left( \pi {-} \tan^{{-}1}(\ell/x) {+} y \right) {-} F_{\varphi|\overline{\varphi}} \left( \pi{-}\tan^{{-}1}(\ell/x) {-} y  \right)$ with $F_{\varphi|\overline{\varphi}}$ being the CDF of $\varphi$ for given $\overline{\varphi}$. In addition, we have $\mathcal{S}_{\overline{\varphi}}(r) {=} \left\lbrace \left[ \overline{\varphi}_1(r),\overline{\varphi}_2(r) \right],\left[ \overline{\varphi}_3(r),\overline{\varphi}_4(r) \right] \right\rbrace$ where $\overline{\varphi}_1(r)\,{=}\,\max(\overline{\varphi}_{\min},\alpha({-}\Theta,r)$, $\overline{\varphi}_2(r)\,{=}\,\min(\overline{\varphi}_{\max},\alpha({-}\theta_{\rm th},r))$, $\overline{\varphi}_3(r)\,{=}\,\max(\overline{\varphi}_{\min},\alpha(\theta_{\rm th},r))$, and $\overline{\varphi}_4(r)\,{=}\,\min(\overline{\varphi}_{\max},\alpha(\Theta,r))$ with $\alpha(x,r) \,{=}\, \pi {-} \tan^{{-}1}(\ell/r) {+} x$. Moreover, $A(x)$ is defined as
\begin{align}\label{eqn:auxiliary_a}
A(x)\,{=}\,I(\Theta,x,d_{\max}) \,{-}\, I({-}\Theta,x,d_{\max}) \,{-}\, I(\theta_{\rm th},x,d_{\max}) \,{+}\, I({-}\theta_{\rm th},x,d_{\max}),
\end{align}
where $I(x,y,z)$ is given as
\begin{align}\label{eqn:integral}
\!\!\!I(x,y,z) = \begin{cases}
\Delta g\left( x, u \left( r_{\min}(x),y,z \right),z \right) & \textrm{ if } \overline{\varphi}_{\min} {-} \pi \leq x < \xi_{\min}, \\ 
\Delta g\left( x, y, z \right) & \textrm{ if } \overline{\varphi}_{\min}{-}{\displaystyle \frac{\pi}{2}} \leq x < \overline{\varphi}_{\max}{-}\pi , \\
\Delta g\left( x, u \left( r_{\min}(x),y,z \right),u \left( r_{\max}(x),y,z \right) \right) & \\
\qquad \qquad + z {-} u\left( r_{\max}(x),y,z \right)  & \textrm{ if } \overline{\varphi}_{\max}{-}\pi \leq x < \overline{\varphi}_{\min}{-}{\displaystyle \frac{\pi}{2}} , \\
\Delta g\left( x,y,u \left( r_{\max}(x),y,z \right) \right) & \\
\qquad \qquad + z {-} u\left( r_{\max}(x),y,z \right) & \textrm{ if } \xi_{\max} \leq x  < \overline{\varphi}_{\max}{-}{\displaystyle \frac{\pi}{2}}  ,\\
z {-} y & \textrm{ if } x \geq \overline{\varphi}_{\max}{-}{\displaystyle \frac{\pi}{2}}, 
\end{cases}
\end{align} 
with $\xi_{\min} {=}\, \min \left( \overline{\varphi}_{\max}{-}\pi, \overline{\varphi}_{\min}{-}{\displaystyle \frac{\pi}{2}} \right)$, $\xi_{\max} {=}\, \max \left( \overline{\varphi}_{\max}{-}\pi, \overline{\varphi}_{\min}{-}{\displaystyle \frac{\pi}{2}} \right)$, $r_{\min}(x) {=}\, h/\tan \left(x{-}\overline{\varphi}_{\min} \right)$, $r_{\max}(x) {=}\, h/\tan \left(x{-}\overline{\varphi}_{\max} \right)$, $\Delta g ( x,a,b ) \,{=}\, g(x,b) \,{-}\, g(x,a)$, and
\begin{align}\label{eqn:exact_integral}
g(x,y) &= \frac{1}{\Delta\overline{\varphi}} \left[ \left( \pi+x-\overline{\varphi}_{\min} \right) y - \frac{\ell}{2} \log(\ell^2 + y^2) - r \tan^{{-}1} \left( \ell/y \right) \right].
\end{align}

Similarly, the CDF of the nonzero square-channel for user $j\,{\in}\,\mathcal{S}_{\rm S, \overline{\varphi}}$ is
\begin{align}
F_{h_j^2|\overline{h}_j>0}(x) = \frac{B(d^*(x))}{B(d_{\min})} + \frac{1}{\Delta \overline{\varphi}B(d_{\min})} \int_{d_{\min}}^{d^*(x)} \int_{\overline{\varphi}_2(r)}^{\overline{\varphi}_3(r)} \left( 1{-} \Delta F_{\varphi|\overline{\varphi}} \left( r,\Psi(x,r,\Theta) \right) \right)  {\rm d}\overline{\varphi} \, {\rm d}r,\label{eqn:cdf_twobit_mean_j}
\end{align}
where $B(x) \,{=}\, I(\theta_{\rm th},x,d_{\rm th}) \,{-}\, I({-}\theta_{\rm th},x,d_{\rm th})$ with $I(x,y,z)$ given in \eqref{eqn:integral}.
\end{theorem}
\begin{IEEEproof}
See Appendix~\ref{app:cdf_twobit_mean}.
\end{IEEEproof}

Finally, we can compute the desired outage probabilities for user $i$ and user $j$, which are given by \eqref{eqn:outpr_i_3} and \eqref{eqn:outpr_j_3} in Section~\ref{sec:outage_formulation}, respectively, by using the respective nonzero square-channel CDFs given in Theorem~\ref{the:cdf_twobit_ins} and Theorem~\ref{the:cdf_twobit_mean}. In Fig.~\ref{fig:cdf_squarechannel}\subref{fig:cdf_squarechannel_two}, we present numerical results verifying the nonzero square-channel CDFs in Theorem~\ref{the:cdf_twobit_ins} and Theorem~\ref{the:cdf_twobit_mean} with the setting of Fig.~\ref{fig:cdfvertang_pmfknz}. We observe that the use of mean vertical angle information in feedback computation results in more deviation for the stronger user $j$ than that for the weaker $i$ user for this specific setting. 

\vspace{-0.15in}
\section{Numerical Results}\label{sec:results}
In this section, we present numerical results for the the sum-rate performance of the NOMA strategies and feedback schemes considered in Section~\ref{sec:noma}. To this end, we consider Monte Carlo based extensive computer simulations as well as analytical expressions of Section~\ref{sec:outage_analysis}. In this regard, we assume a total of $K\,{=}\,20$ users, each of which has a uniformly distributed horizontal distance within $d_{\min}\,{=}\,0$~m and $d_{\max}\,{=}\,10$~m. Moreover, respective mean vertical angles independently follow uniform distribution within $\overline{\varphi}_\text{min}\,{=}\,\Delta\varphi$ and $\overline{\varphi}_\text{max}\,{=}\,180-\Delta\varphi$, so that the instantaneous vertical angle spans $[0^\circ,180^\circ]$ irrespective of the particular $\Delta\varphi$ value. 

We also assume that the LED is vertically off the horizontal plane by $\ell\,{=}\,2$~m with a half power beamwidth of $\Phi_{\rm HPBW}\,{=}\,60^\circ$, and that the receiver area of the photodetector is $A_{\rm e}\,{=}\,1~\text{cm}^2$. We choose the power allocation coefficients to be $\beta_i\,{=}\,63/64$ and $\beta_j\,{=}\,1/64${\color{black}, which is in agreement with the power allocation strategy described in Section~\ref{sec:noma_userrates}.} {\color{black}The respective target data rates for $i$th and $j$th users} are assumed to be $\overline{R}_i\,{=}\,2$~\mbox{bit/s/Hz} and $\overline{R}_j\,{=}\,10$~\mbox{bit/s/Hz}{\color{black}, respectively}. {\color{black}Finally, we assume two different values for the FOV of the photodetectors, which are $100^\circ$ and $180^\circ$ (i.e., $\Theta\,{=}\,\{50^\circ,90^\circ\}$) being representative of a relatively narrow and wide FOV, respectively}.

\vspace{-0.15in}
\subsection{Sum-Rate Performance for Individual User Scheduling}\label{sec:numerical_individual}
\begin{figure}[!t]
\centering
\subfloat[{\color{black} $\text{FOV}\,{=}\,100^\circ$ ($\Theta\,{=}\,50^\circ$)} ]{\includegraphics[width=0.5\textwidth]{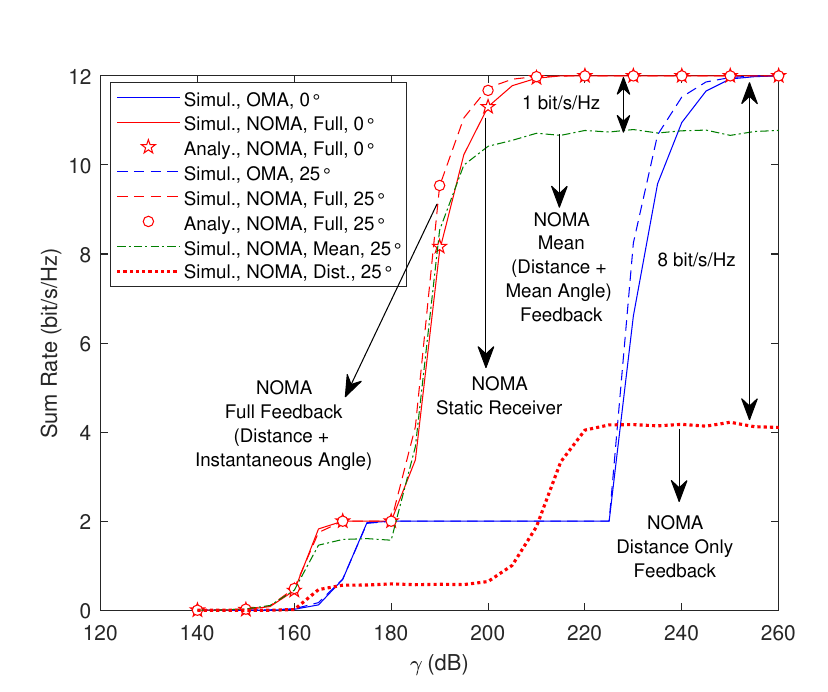}
\label{fig:nogroup_sumrate_fov100}}
\subfloat[{\color{black} $\text{FOV}\,{=}\,180^\circ$ ($\Theta\,{=}\,90^\circ$)} ]{\includegraphics[width=0.5\textwidth]{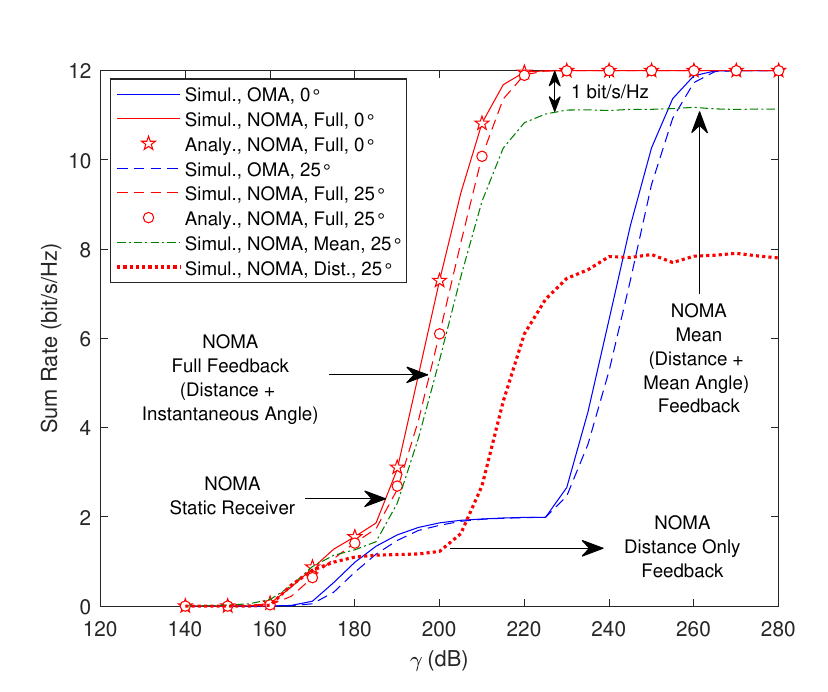}
\label{fig:nogroup_sumrate_fov180}}
\caption{OMA and NOMA sum rates against transmit SNR ($\gamma$) for individual user scheduling strategy of Section~\ref{sec:noma_feedback_individual}, where $\Delta\varphi\,{=}\,\{0^\circ,25^\circ\}$ and $\text{\color{black}FOV}\,{=}\,\{100^\circ,180^\circ\}$ {\color{black}(i.e., $\Theta\,{=}\,\{50^\circ,90^\circ\}$)}. }
\label{fig:nogroup_sumrate}\vspace{-0.3in}
\end{figure}

We start with the sum-rate performance of NOMA for the individual user scheduling strategy of Section~\ref{sec:noma_feedback_individual}. In Fig.~\ref{fig:nogroup_sumrate}, we depict the sum rates of OMA and NOMA for the user indices $i\,{=}\,1$ and $j\,{=}\,10$ with the maximum vertical deviation angle $\Delta\varphi\,{=}\,\{0^\circ,25^\circ\}$. Note that while $\Delta\varphi\,{=}\,0^\circ$ corresponds to ``\textit{static}'' receiver orientation in the vertical domain, $\Delta\varphi\,{=}\,25^\circ$ represents ``\textit{dynamic}'' receiver orientation with a large variation (i.e., orientation spans as large as $50^\circ$ in the vertical domain over time). We observe that NOMA outperforms OMA in terms sum rates, and that analytical results nicely follow the simulation data in all the cases. 

We note that the performance of the mean vertical angle based limited feedback is very close to that of the full CSI based feedback even for a large $\Delta\varphi$ value of $25^\circ$. The respective sum-rate degradation at the steady state is around $1$~\mbox{bit/s/Hz} only for both FOV choices. On the other hand, the distance {\color{black}only} feedback cannot capture the FOV status (i.e., whether the receive direction is inside the FOV or not) correctly when $\Delta\varphi$ gets larger, and the steady-state sum-rate loss can therefore be as large as $8$~\mbox{bit/s/Hz}, as shown in Fig.~\ref{fig:nogroup_sumrate}\subref{fig:nogroup_sumrate_fov100}. Note also that any change in vertical orientation is likely to alter the FOV status of more users as we have narrower FOV. As a result, the distance feedback scheme represents the FOV status very poorly along with increasing $\Delta\varphi$ when the FOV is narrow. This is the reason for the steady-state sum-rate degradation of the distance feedback in Fig.~\ref{fig:nogroup_sumrate}\subref{fig:nogroup_sumrate_fov100} ({\color{black}FOV is} $100^\circ$) as compared to Fig.~\ref{fig:nogroup_sumrate}\subref{fig:nogroup_sumrate_fov180} ({\color{black}FOV is} $180^\circ$).

\begin{remark}\label{rem:indiv}
Note that the full CSI based feedback results in a better sum-rate performance in the transition region (i.e., $\gamma\,{\leq}\,220$) for the dynamic case (i.e., with $\Delta\varphi\,{=}\,25^\circ$) as compared to the static case (i.e., with $\Delta\varphi\,{=}\,0^\circ$), when the FOV is relatively narrow, as shown in Fig.~\ref{fig:nogroup_sumrate}\subref{fig:nogroup_sumrate_fov100}. Similar to the previous discussion, a narrower FOV is likely to result in more users having zero channel gain since the receive direction can easily fall outside the FOV. On the other hand, as the vertical orientation spans wider angular domain with increasing $\Delta\varphi$, some of these users having zero channel gain initially can establish a direct link with the LED such that their receive directions fall inside this narrow FOV again. This is the reason for the improvement of the NOMA sum rates with full CSI feedback as $\Delta\varphi$ increases from $0^\circ$ to $25^\circ$, as shown in Fig.~\ref{fig:nogroup_sumrate}\subref{fig:nogroup_sumrate_fov100}. 
\end{remark}

\begin{figure}[!t]
\centering
\includegraphics[width=0.6\textwidth]{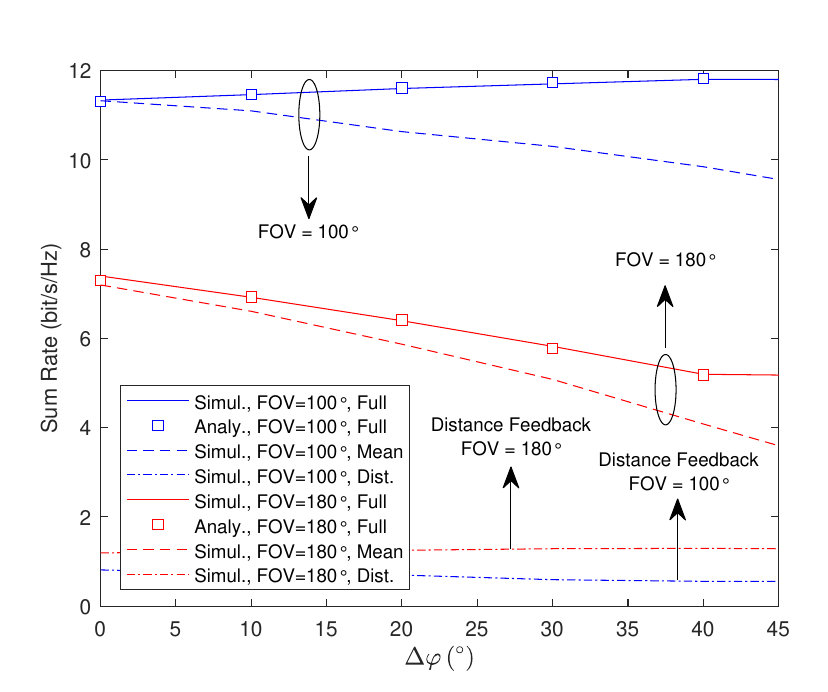}\vspace{-0.3in}
\caption{NOMA sum rates against $\Delta\varphi$ for individual user scheduling strategy of Section~\ref{sec:noma_feedback_individual}, where $\text{\color{black}FOV}\,{=}\,\{100^\circ,180^\circ\}$ {\color{black}(i.e., $\Theta\,{=}\,\{50^\circ,90^\circ\}$)} and transmit SNR is $\gamma\,{=}200\,\text{dB}$.}
\label{fig:nogroup_sumrate_vardev}\vspace{-0.25in}
\end{figure}

In Fig.~\ref{fig:nogroup_sumrate_vardev}, we depict the sum-rate performance of NOMA with varying maximum deviation angle $\Delta\varphi$ for a transmit SNR of $200\,\text{dB}$, and user indices of $i\,{=}\,1$ and $j\,{=}\,10$. As before, analytical results successfully match the simulation data even for large $\Delta\varphi$ values (e.g., $\Delta\varphi \,{=}\, 45^\circ$ indicates a span of $90^\circ$ in the vertical domain). We observe that the performance of the mean vertical angle based feedback deviates from that of the full CSI based feedback with increasing $\Delta\varphi$ for both FOV choices. As $\Delta\varphi$ increases, the sum-rate performance of the full CSI feedback improves for {\color{black}a narrower FOV of} $100^\circ$, and degrades for {\color{black}a wider FOV of} $180^\circ$, which aligns with the discussion of Remark~\ref{rem:indiv}. Note that although the full CSI feedback with large $\Delta\varphi$ can have a better performance as compared to the static case, this feedback scheme is impractical due to the necessity of the continuous tracking and feedback of the variation in the vertical angle. 

\vspace{-0.15in}
\subsection{Sum-Rate Performance for Group-Based User Scheduling}\label{sec:numerical_group}

\begin{figure}[!t]
\centering
\subfloat[{\color{black} $\text{FOV}\,{=}\,100^\circ$ ($\Theta\,{=}\,50^\circ$)} ]{\includegraphics[width=0.5\textwidth]{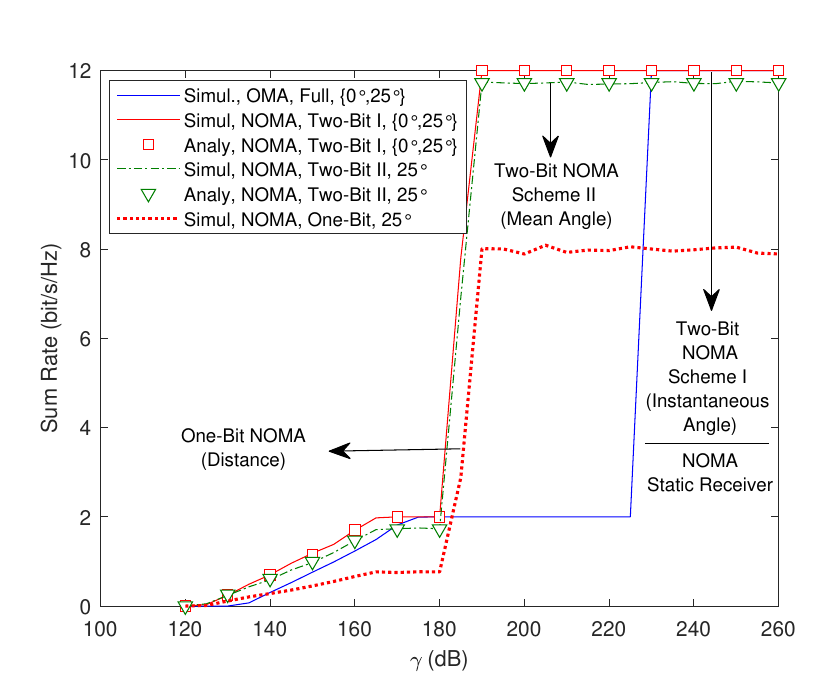}
\label{fig:twobit_sumrate_fov100}}
\subfloat[{\color{black} $\text{FOV}\,{=}\,180^\circ$ ($\Theta\,{=}\,90^\circ$)} ]{\includegraphics[width=0.5\textwidth]{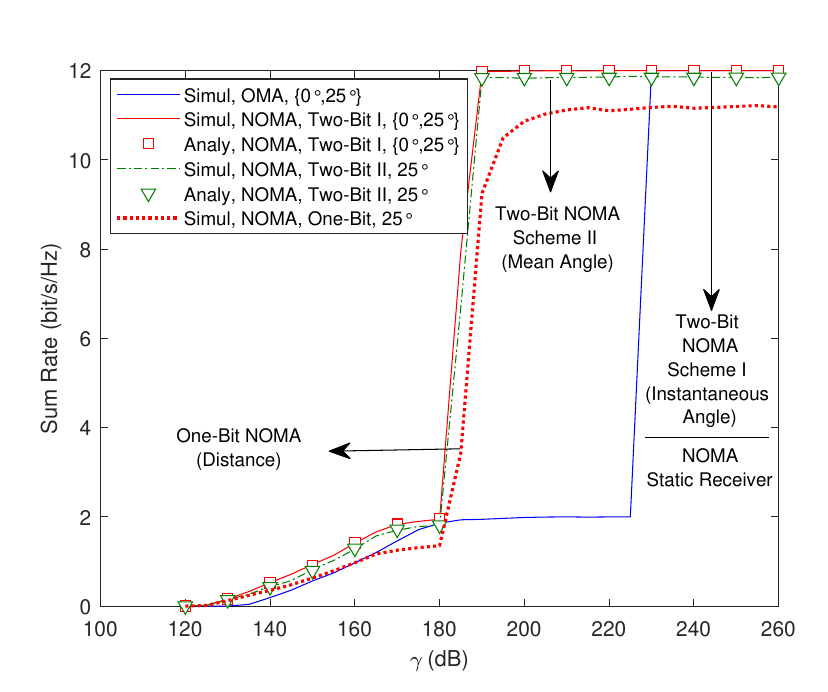}
\label{fig:twobit_sumrate_fov18}}
\caption{OMA and NOMA sum rates against transmit SNR ($\gamma$) for group-based user scheduling with $\Delta\varphi\,{=}\,\{0^\circ,25^\circ\}$, $\text{\color{black}FOV}\,{=}\,\{100^\circ,180^\circ\}$ {\color{black}(i.e., $\Theta\,{=}\,\{50^\circ,90^\circ\}$)}, $d_{\rm th} {=}\, 1\,\text{m}$ $\left( c_{d_{\rm th}} {=}\, 0.1 \right)$, and $\theta_{\rm th} {=}\,5^\circ$ $\left( c_{\theta_{\rm th}} {=}\, 0.1 \right)$.}
\label{fig:twobit_sumrate}\vspace{-0.3in}
\end{figure}

We now consider the sum-rate performance of NOMA with the group-based user scheduling strategy of Section~\ref{sec:noma_feedback_group} with two-bit feedback. Without any loss of generality, we assume that $d_{\rm th} \,{=}\, d_{\min} \,{+}\, c_{d_{\rm th}} \left( d_{\max} \,{-}\, d_{\min} \right)$ and $\theta_{\rm th} \,{=}\, c_{\theta_{\rm th}} \Theta$, where $c_{d_{\rm th}} {\in}\, [0,1]$ and $c_{\theta_{\rm th}} {\in}\, [0,1]$ are the threshold coefficients to determine the values of $d_{\rm th}$ and $\theta_{\rm th}$, respectively. In Fig.~\ref{fig:twobit_sumrate}, we plot the sum-rate results of OMA and NOMA with varying transmit SNR, where $c_{d_{\rm th}} \,{=}\, 0.1$, $c_{\theta_{\rm th}} \,{=}\, 0.1$, and $\Delta\varphi\,{=}\,\{0^\circ,25^\circ\}$. As before, the analytical results nicely matches the experimental data, and NOMA achieves better sum-rate performance than OMA.

In Fig.~\ref{fig:twobit_sumrate}, we also observe that the sum-rate performance for NOMA with two-bit feedback of the distance $d$ and the instantaneous vertical angle $\varphi$, {\color{black}which is referred to as Scheme I in Fig.~\ref{fig:twobit_sumrate} for the sake of clarity}, remains the same as $\Delta\varphi$ increases from $0^\circ$ (i.e., static receiver orientation) to $25^\circ$ (i.e., dynamic receiver orientation), regardless of the FOV. {\color{black}\textit{Two-bit feedback based NOMA is therefore very robust to the random receiver orientation}}. When the mean vertical angle $\overline{\varphi}$ is employed instead of $\varphi$ in the feedback computation (together with the distance), {\color{black}which is referred to as Scheme II in Fig.~\ref{fig:twobit_sumrate}}, the degradation in NOMA sum rates is less than $0.3$~\mbox{bit/s/Hz} at the steady state for both FOV choices. {\color{black}\textit{This result underscores the power of the practical feedback scheme for two-bit feedback based NOMA involving mean vertical angle (instead of using its instantaneous value)}}. Moreover, the one-bit feedback involving only the distance $d$ (i.e., without vertical angle) results in much worse sum-rate performance than that of the previous two-bit feedback strategies, with a loss of as large as $4$~\mbox{bit/s/Hz} at the steady state for the FOV of $100^\circ$. 

\begin{remark}\label{rem:threshold_coeff}
Note that whenever $c_{d_{\rm th}}$ or $c_{\theta_{\rm th}}$ gets smaller values, i) the set $\mathcal{S}_{\rm S, \varphi}$ ($\mathcal{S}_{\rm S, \overline{\varphi}}$) has fewer users, and ii) the set $\mathcal{S}_{\rm W, \varphi}$ ($\mathcal{S}_{\rm W, \overline{\varphi}}$) has more users, where the respective channel gains {\color{black}in each group} increase on the average. Although stronger users {\color{black}in $\mathcal{S}_{\rm S, \varphi}$ ($\mathcal{S}_{\rm S, \overline{\varphi}}$) are desired for better user rates}, we need to have {\color{black}\textit{sufficiently} weaker users} in $\mathcal{S}_{\rm W, \varphi}$ ($\mathcal{S}_{\rm W, \overline{\varphi}}$) to keep the strong and weak NOMA users \textit{more distinctive} in the power domain, {\color{black}and therefore to achieve} a better decoding performance. As a result, the choice of the threshold coefficients $c_{d_{\rm th}}$ and $c_{\theta_{\rm th}}$ impacts the respective sum-rate performance, and therefore necessitates the formulation of an optimization problem. Note also that $c_{d_{\rm th}}$ and $c_{\theta_{\rm th}}$ affects the number of users in each of these sets, and some particular threshold choices (i.e., very small or very large) may end up with either $\mathcal{S}_{\rm S, \varphi}$ ($\mathcal{S}_{\rm S, \overline{\varphi}}$) or $\mathcal{S}_{\rm W, \varphi}$ ($\mathcal{S}_{\rm W, \overline{\varphi}}$) being an empty set for which the NOMA transmission is impossible. Hence, the optimization of $c_{d_{\rm th}}$ and $c_{\theta_{\rm th}}$ needs to consider the alternative \textit{single-user} transmission techniques whenever NOMA is not viable {\color{black}(see \cite{Yapici2018AngFee, Yapici2018NOMAmmWDro_J} for details)}.
\end{remark}

\begin{figure}[!t]
\centering
\subfloat[{\color{black} $\text{FOV}\,{=}\,100^\circ$ ($\Theta\,{=}\,50^\circ$)} ]{\includegraphics[width=0.5\textwidth]{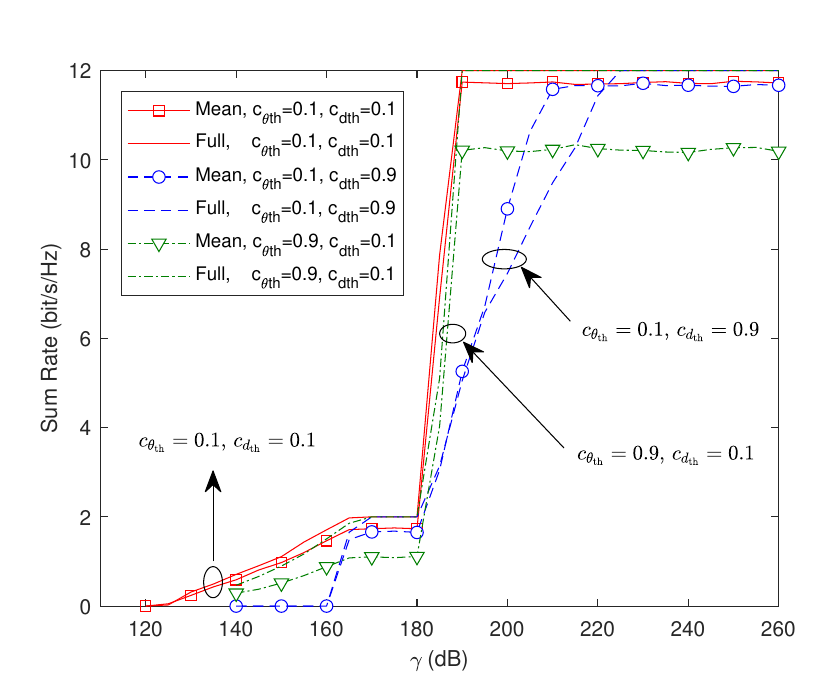}
\label{fig:twobit_sumrate_fov100_cangvar_cdisvar_dev25}}
\subfloat[{\color{black} $\text{FOV}\,{=}\,180^\circ$ ($\Theta\,{=}\,90^\circ$)} ]{\includegraphics[width=0.5\textwidth]{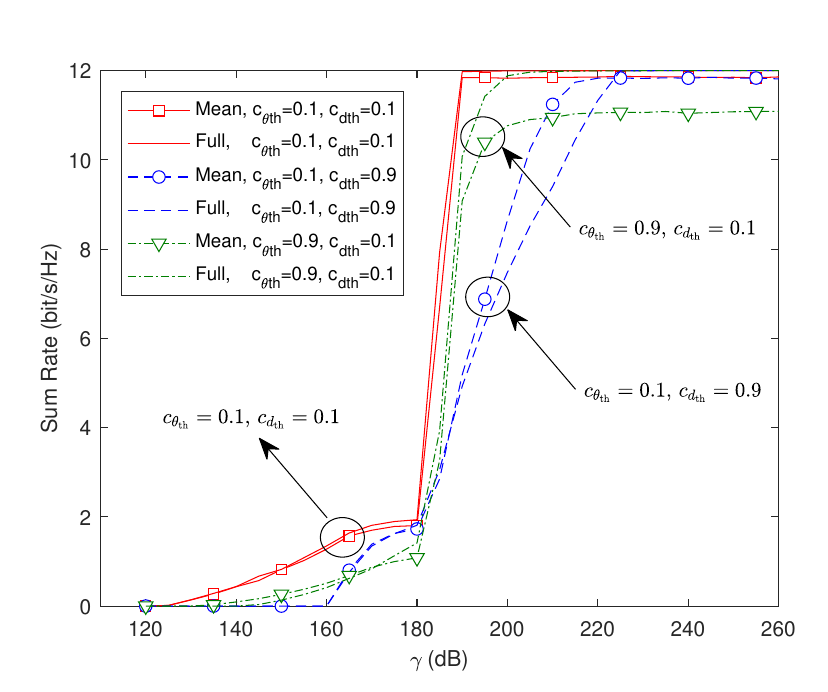}
\label{fig:twobit_sumrate_fov180_cangvar_cdisvar_dev25}}
\caption{\color{black}Simulation results for NOMA sum rates against transmit SNR ($\gamma$) for group-based user scheduling with $\Delta\varphi\,{=}\,25^\circ$, $\text{FOV}\,{=}\,\{100^\circ,180^\circ\}$ (i.e., $\Theta\,{=}\,\{50^\circ,90^\circ\}$), and $c_{d_{\rm th}},\,c_{\theta_{\rm th}} {\in}\, \{0.1,0.9\}$. The NOMA feedback mechanism is two-bit involving the distance as well as either the instantaneous (i.e., full) or mean value of the vertical angle.}
\label{fig:twobit_sumrate_cangvar_cdisvar_dev25}\vspace{-0.3in}
\end{figure}

{\color{black}To gain some more insight into the proper selection of threshold coefficients, we depict the sum-rate performance of two-bit feedback NOMA in Fig.~\ref{fig:twobit_sumrate_cangvar_cdisvar_dev25} for the dynamic receiver orientation scenario with $\Delta\varphi\,{=}\,25^\circ$.  We observe for this particular scenario that the threshold values of $c_{d_{\rm th}}{=}\,c_{\theta_{\rm th}} {=}\, 0.1$ work the best for both FOV values and any choice of transmit SNR. We also note that as the angle threshold gets larger (e.g., $c_{\theta_{\rm th}}{=}\, 0.9$), the steady-state sum-rate performance of two-bit feedback involving mean vertical angle $\overline{\varphi}$ deteriorates since, in part, the FOV status cannot be captured accurately through $\overline{\varphi}$ any more. On the other hand, larger distance threshold (e.g., $c_{d_{\rm th}}{=}\, 0.9$) necessitates more transmit power to reach to the steady-state performance.} 

\begin{remark}\label{rem:}
The performance of the NOMA strategy with individual user scheduling evaluated in Section~\ref{sec:numerical_individual} does not consider any optimization over the stronger and weaker user indices, which is basically due to the reasoning in Remark~\ref{rem:threshold_coeff}. We therefore do not provide any comparison between the individual and grouped-based user scheduling NOMA strategies, since none of them is optimized to yield its best performance in terms of specific user choices. Moreover, the steady-state sum-rate performance of the feedback mechanism relying on $\overline{\varphi}$ is observed to be much closer to that of $\varphi$, when the group-based user scheduling with two-bit feedback is adopted. Furthermore, the feedback mechanism relying only on $d$ has always worse sum-rate performance in either scheduling strategy, which deteriorates even more together with narrower FOV values.
\end{remark}

\begin{figure}[!t]
\centering
\includegraphics[width=0.6\textwidth]{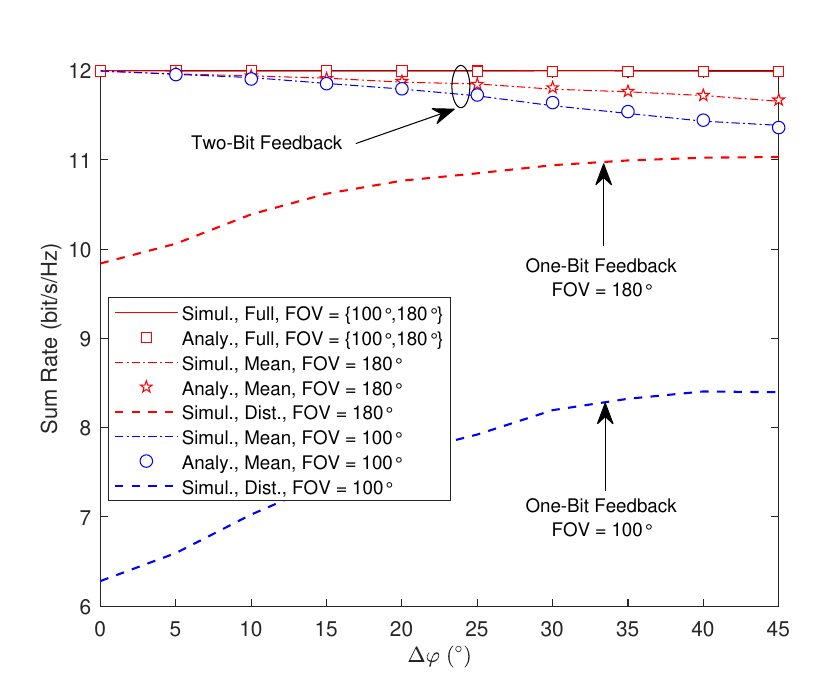}\vspace{-0.3in}
\caption{NOMA sum rates against $\Delta\varphi$ for group-based user scheduling of Section~\ref{sec:noma_feedback_group}, where $\text{\color{black}FOV}\,{=}\,\{100^\circ,180^\circ\}$ {\color{black}(i.e., $\Theta\,{=}\,\{50^\circ,90^\circ\}$)}, $d_{\rm th} {=}\, 1$~m $\left( c_{d_{\rm th}} {=}\, 0.1 \right)$, $\theta_{\rm th} {=}\,5^\circ$ $\left( c_{\theta_{\rm th}} {=}\, 0.1 \right)$, and transmit SNR is $\gamma\,{=}200$~dB.}
\label{fig:twobit_sumrate_fovvar_cang01_cdis01_devvar}\vspace{-0.25in}
\end{figure}

In Fig.~\ref{fig:twobit_sumrate_fovvar_cang01_cdis01_devvar}, we demonstrate the sum-rate performance of NOMA along with varying maximum deviation angle $\Delta\varphi$ for $c_{d_{\rm th}} \,{=}\, 0.1$, $c_{\theta_{\rm th}} \,{=}\, 0.1$, and transmit SNR of $200$~dB. We observe that the analytical results successfully match the simulation data in all the cases under consideration. Note that the performance of two-bit feedback with $d$ and $\overline{\varphi}$ degrades along with increasing $\Delta\varphi${\color{black}, since the mean value $\overline{\varphi}$ starts becoming poor representative of its instantaneous value $\varphi$}. Moreover, the respective sum-rate loss turns out to be larger for the narrower FOV of $100^\circ$, since $\overline{\varphi}$ captures the FOV status hardly for relatively narrower FOV values as $\Delta\varphi$ becomes larger. Nevertheless, the practical two-bit feedback scheme with $d$ and $\overline{\varphi}$ is superior to the one-bit feedback with $d$ only{\color{black}, and performs very close to the optimal two-bit feedback with $d$ and $\varphi$} for any $\Delta\varphi$ choice.

\subsection{\color{black}Impact of Noisy Distance and Angle Information}\label{sec:numerical_noisy}

\begin{figure}[!t]
\centering
\subfloat[{\color{black} $\text{FOV}\,{=}\,100^\circ$ ($\Theta\,{=}\,50^\circ$)} ]{\includegraphics[width=0.5\textwidth]{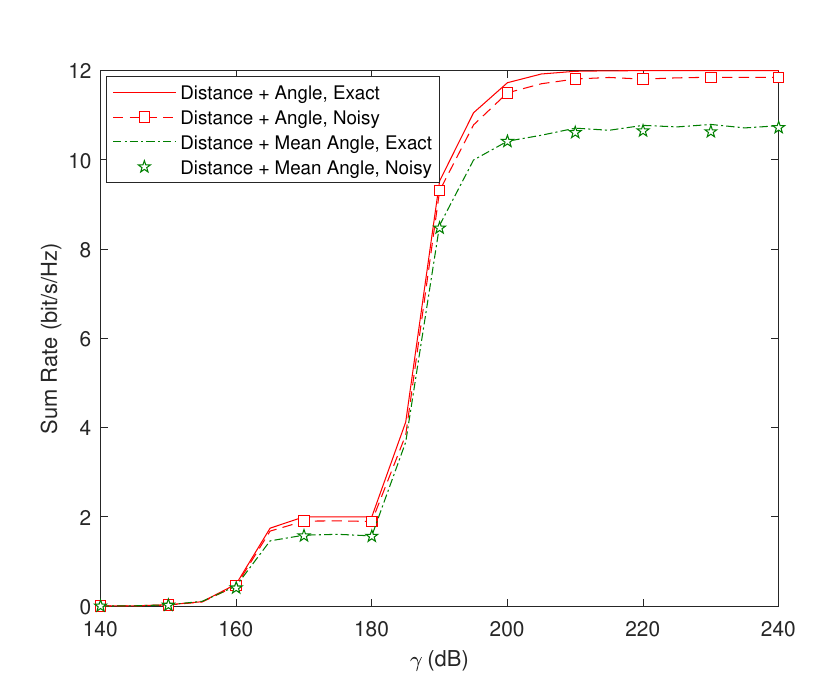}}
\subfloat[{\color{black} $\text{FOV}\,{=}\,180^\circ$ ($\Theta\,{=}\,90^\circ$)} ]{\includegraphics[width=0.5\textwidth]{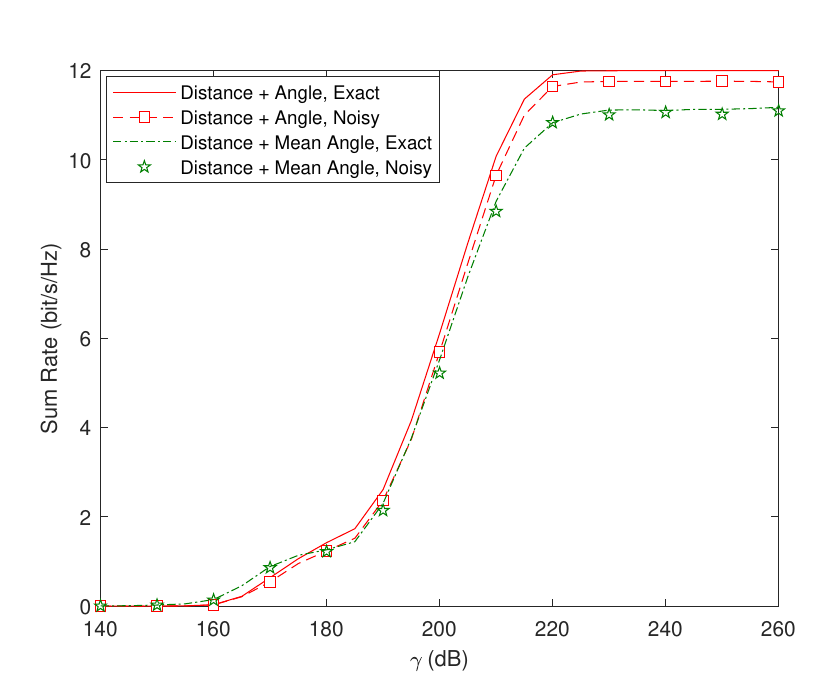}}
\caption{\color{black}Simulation results for NOMA sum rates against transmit SNR ($\gamma$) for individual user scheduling strategy, where distance and angle error for noisy case is $0.1\,\text{m}$ and $5^\circ$, respectively,    $\Delta\varphi\,{=}\,25^\circ$ and $\text{FOV}\,{=}\,\{100^\circ,180^\circ\}$ (i.e., $\Theta\,{=}\,\{50^\circ,90^\circ\}$). }
\label{fig:nogroup_sumrate_noisy}
\vspace{-0.3in}
\end{figure}

{\color{black}We finally consider the impact of \textit{noisy} horizontal distance and vertical angle information on the NOMA sum rates in a \textit{dynamic} scenario with $\Delta\varphi\,{=}\,25^\circ$. In order to represent any imperfections regarding the limited-feedback information, we consider noisy estimates of distance and angle given as $\hat{d}_k \,{=}\, d_k \,{+}\, \epsilon_d$, $\hat{\varphi}_k \,{=}\, \varphi_k \,{+}\, \epsilon_\varphi$, and $\hat{\overline{\varphi}}_k \,{=}\, \overline{\varphi}_k \,{+}\, \epsilon_\varphi$. In this representation, $\epsilon_d$ and $\epsilon_\varphi$ stand for estimation error in distance and angle, respectively, and are assumed to be complex Gaussian with zero-mean and variance $\sigma_d^2$ and $\sigma_\varphi^2$, respectively. We assume that $\sigma_d \,{=}\, 0.05$ and $\sigma_\varphi \,{=}\, 2.5$, which correspond to  $0.1\,\text{m}$ error in distance and $5^\circ$ error in vertical angle. Note that these error values can be achieved through various received signal strength (RSS) based localization techniques including the one in~\cite{Lau2018SimPos}, which can also be applied during data transmission phase, as well. In Fig.~\ref{fig:nogroup_sumrate_noisy}, we depict the simulation results for individual user-scheduling NOMA with noisy distance and angle information assuming the same setting of Fig.~\ref{fig:nogroup_sumrate}. We observe that the performance of the mean vertical angle based limited-feedback scheme does not change while that of the instantaneous angle based scheme exhibits only a marginal degradation.}

\vspace{0.0in}
\section{Conclusion}\label{sec:conclusion}
We investigated a downlink multiuser VLC scenario involving mobile users with random vertical orientation. In order to increase the spectral efficiency, the NOMA transmission is employed with various user scheduling techniques and feedback mechanisms. The outage probability and sum-rate expressions are derived analytically, where the respective numerical results show a very good match with the simulation data. We observe that the practical feedback scheme with the mean vertical angle achieves a near-optimal sum-rate performance. In addition, the two-bit feedback involving both the distance and the angle information significantly outperforms the conventional one-bit feedback with the distance information only.

\appendices

\vspace{-0.2in}
\section{Proof of Lemma~\ref{lem:cdf_varphi}}\label{app:cdf_varphi_derivation}
The desired CDF of $\varphi$ is given as
\begin{align}\label{app:cdf_varphi}
F_{\varphi}(x) &= \Pr \left( \varphi \leq x \right) = \frac{1}{\Delta\overline{\varphi}}\int_{\overline{\varphi}_{\rm min}}^{\overline{\varphi}_{\rm max}} \Pr \left( \varphi \leq x | \overline{\varphi} \right) {\rm d}\overline{\varphi} ,
\end{align}
where
\begin{align}\label{app:cdf_varphi_cond}
\Pr \left( \varphi \leq x | \overline{\varphi} \right) = \begin{cases}
1 & \textrm{ if } x \geq \overline{\varphi} + \Delta\varphi,\\
\displaystyle \frac{x - \overline{\varphi} + \Delta\varphi}{2\Delta\varphi} &  \textrm{ if } \overline{\varphi}-\Delta\varphi \leq x < \overline{\varphi}+\Delta\varphi ,\\
0 &  \textrm{ if } x < \overline{\varphi} - \Delta\varphi .
\end{cases}
\end{align}
In order to compute $F_{\varphi}(x)$, we need to consider the integration in \eqref{app:cdf_varphi} over $\overline{\varphi} \,{\in}\, [\overline{\varphi}_{\rm min},\overline{\varphi}_{\rm max}]$, and the intervals $\overline{\varphi} \,{\in}\, (x{-}\Delta\varphi,x{+}\Delta\varphi]$ and $\overline{\varphi} \,{\in}\, ({-}\infty ,x{-}\Delta\varphi]$ of \eqref{app:cdf_varphi_cond} jointly. We readily have $F_{\varphi}(x) \,{=}\, 1$ for $x \,{\geq}\, \overline{\varphi}_{\rm max} \,{+}\, \Delta\varphi$, and $F_{\varphi}(x) \,{=}\, 0$ for $x \,{<}\, \overline{\varphi}_{\rm min} \,{-}\, \Delta\varphi$, and the following possibilities.

\paragraph{Case 1} $x \,{-}\, \Delta\varphi \,{\leq}\, \overline{\varphi}_{\rm min} \,{\leq}\, x \,{+}\, \Delta\varphi \,{\leq}\, \overline{\varphi}_{\rm max} $
\begin{align}
F_{\varphi}(x) &= \frac{1}{2\Delta\varphi \, \Delta\overline{\varphi}} \int_{\overline{\varphi}_{\rm min}}^{x \,{+}\, \Delta\varphi } \left( x - \overline{\varphi} + \Delta\varphi \right) {\rm d}\overline{\varphi} = \frac{\left(x+\Delta\varphi - \overline{\varphi}_{\rm min}\right)^2}{4 \Delta\varphi \Delta\overline{\varphi}}.
\end{align}

\paragraph{Case 2} $\overline{\varphi}_{\rm min} \,{\leq}\, x \,{-}\, \Delta\varphi  \,{\leq}\, \overline{\varphi}_{\rm max} \,{\leq}\, x \,{+}\, \Delta\varphi $ 
\begin{align}
F_{\varphi}(x) &= \frac{1}{\Delta\overline{\varphi}} \left[ \int_{\overline{\varphi}_{\rm min}}^{x \,{-}\, \Delta\varphi } {\rm d}\overline{\varphi} + \frac{1}{2\Delta\varphi} \int_{x \,{-}\, \Delta\varphi }^{\overline{\varphi}_{\rm max}}  \left( x - \overline{\varphi} + \Delta\varphi \right) {\rm d}\overline{\varphi} \right] = 1-\frac{\left( \overline{\varphi}_{\rm max} - x + \Delta\varphi \right)^2}{4 \Delta\varphi \Delta\overline{\varphi}}.\label{eqn:case2}
\end{align}

\paragraph{Case 3} $\overline{\varphi}_{\rm min} \,{\leq}\, x \,{-}\, \Delta\varphi \,{\leq}\, x \,{+}\, \Delta\varphi \,{\leq}\, \overline{\varphi}_{\rm max} $
\begin{align}
F_{\varphi}(x) &= \frac{1}{\Delta\overline{\varphi}} \left[ \int_{\overline{\varphi}_{\rm min}}^{x \,{-}\, \Delta\varphi } {\rm d}\overline{\varphi} + \frac{1}{2\Delta\varphi} \int_{x \,{-}\, \Delta\varphi }^{x \,{+}\, \Delta\varphi }  \left( x - \overline{\varphi} + \Delta\varphi \right) {\rm d}\overline{\varphi} \right] = \frac{x-\overline{\varphi}_{\rm min}}{\Delta\overline{\varphi}}.
\end{align}

\paragraph{Case 4} $x \,{-}\, \Delta\varphi \,{\leq}\, \overline{\varphi}_{\rm min} \,{\leq}\, \overline{\varphi}_{\rm max} \,{\leq}\, x \,{+}\, \Delta\varphi  $
\begin{align}
F_{\varphi}(x) &= \frac{1}{2\Delta\varphi \, \Delta\overline{\varphi}} \int_{\overline{\varphi}_{\rm min}}^{\overline{\varphi}_{\rm max}} \left( x - \overline{\varphi} + \Delta\varphi \right) {\rm d}\overline{\varphi} = \frac{1}{2\Delta\varphi}\left( x+\Delta\varphi-\frac{\overline{\varphi}_{\rm min}+\overline{\varphi}_{\rm max}}{2} \right).
\end{align}
When we employ the expressions for $F_{\varphi}(x)$ in Case $1{-}4$, and rearrange the respective inequalities as a function of $x$, we obtain \eqref{eqn:cdf_varphi}. \hfill\IEEEQEDhere

\vspace{-0.2in}
\section{Proof of Lemma~\ref{lem:numof_nzusers}}\label{app:numof_nzusers}
Let $\textsc{I}_{\Theta}(\theta_k)$ be an indicator function defined as $\textsc{I}_{\Theta}(\theta_k) \,{=}\, \Pi \big[ |\theta_k| / \Theta \big]$, where $\Pi$ function is given along with \eqref{eqn:channel}. We therefore have $\textsc{I}_{\Theta}(\theta_k) \,{=}\,1$ and $\textsc{I}_{\Theta}(\theta_k) \,{=}\,0$ representing nonzero and zero channel gains, respectively, and subscript $k$ denoting an arbitrary \textit{unordered} user $k$. Note that, any contribution of the term $\cos^m( \phi_k )$ in \eqref{eqn:channel} to zero channel status is circumvented by a reasonable choice of FOV as $0\,{\leq}\,\Theta\,{<}\,\pi/2$, as discussed in Appendix~\ref{app:cdf_fullcsi}. The zero channel gain is therefore only due to the FOV status (i.e., whether receiver direction is within the FOV). 

Note that $\textsc{I}_{\Theta}(\theta_k) \,{=}\,1$ can be characterized by a probability assignment $p \,{=}\, \Pr \left( |\theta_k| \leq \Theta\right)$, and $\textsc{I}_{\Theta}(\theta_k)$ is therefore a Bernoulli random variable with the event probabilities $p$ and $1{-}p$. In addition, $K_{\rm nz}$ can be represented as a sum of all the indicator functions, which is given as
\begin{align}
K_{\rm nz} = \sum\limits_{k=1}^{K} \textsc{I}_{\Theta}(\theta_k).
\end{align}
Since the sum of Bernoulli random variables defines a Binomial process, $K_{\rm nz}$ becomes a Binomial random variable with the number of independent trials $K$ and the success probability $p$. Note that the success probability is actually equal to the CDF of the absolute value of the incidence angle evaluated at $|\theta_k| \,{=}\,\Theta$, and can therefore be derived as
\begin{align}
p &= \Pr \left( \left|\theta_k \right| \leq \Theta\right) = \Pr \left( \left| \pi - \tan^{{-}1}(\ell/d_k) - \varphi_k \right|\leq \Theta\right) ,\label{eqn:cdf_incidence_1}\\
&= \Pr \left( \pi - \Theta \leq \tan^{{-}1}(\ell/d_k) + \varphi_k \leq \pi + \Theta \right) ,\label{eqn:cdf_incidence_2}\\
&= \frac{1}{\Delta d}\int_{d_{\rm min}}^{d_{\rm max}} \Pr \left( \pi - \Theta - \tan^{{-}1}(\ell/r) \leq \varphi_k \leq \pi + \Theta + \tan^{{-}1}(\ell/r)\right) {\rm d}r,\label{eqn:cdf_incidence_3}
\end{align}
where the last line is obtained by averaging \eqref{eqn:cdf_incidence_2} over the distribution of $d$ being uniform with $\mathcal{U}\,[d_{\rm min}{,}\,d_{\rm max}]$. Expressing \eqref{eqn:cdf_incidence_3} in terms of the CDF of $\varphi$ yields \eqref{eqn:success_prob}. \hfill\IEEEQEDhere

\vspace{-0.2in}
\section{Proof of Theorem~\ref{the:cdf_unordered}}\label{app:cdf_fullcsi}
The desired CDF of the unordered nonzero square-channel is given as
\begin{align} \label{app:cdf_unordered_0}
F_{h^2|h>0}(x) = \Pr \left( h^2 \leq x \,\big| h > 0 \right) ,
\end{align}
where the square-channel is obtained by employing \eqref{eqn:incidenceang} in \eqref{eqn:channel} as follows
\begin{align}\label{app:square_channel}
h^2 = \frac{\cos^2 \left( \beta(\varphi,d) \right)}{\upsilon(d)} \, \Pi\big[ \left| \pi - \beta(\varphi,d) \big| /\Theta \right],
\end{align}
where $\beta(\varphi,d) \,{=}\, \tan^{{-}1}(\ell/d) + \varphi$. Note that the nonzero channel gain may either be due to cosine term or $\Pi$ function in \eqref{app:square_channel}, and, hence, $\varphi$ and $d$ values for nonzero channel gain are
\begin{align}\label{app:nonzero_condition}
\left\lbrace \varphi,d \,\Big|\, \beta(\varphi,d) \neq \left\lbrace \frac{\pi}{2},\frac{3\pi}{2} \right\rbrace, \big| \pi - \beta(\varphi,d) \big| \leq \Theta \right\rbrace .
\end{align}
In \eqref{app:nonzero_condition}, we use the geometrical relations $0\,{\leq}\,\tan^{{-}1}(\ell/d)\,{<}\,\pi/2$ and $0\,{\leq}\,\varphi\,{\leq}\,\pi$, which can be jointly expressed as $0\,{\leq}\, \beta(\varphi,d) \,{<}\,3\pi/2$. Note that we can safely assume $0\,{\leq}\,\Theta\,{<}\,\pi/2$ considering the available LED technology on the market. It is therefore sufficient to consider only $\left| \pi {-} \beta(\varphi,d) \right| \leq \Theta$ to make sure that the channel gain is nonzero.

Hence, the desired CDF in \eqref{app:cdf_unordered_0} becomes
\begin{align} \label{app:cdf_unordered_1}
F_{h^2|h>0}(x) = \frac{\Pr \left( h^2 \leq x , \left| \pi {-} \beta(\varphi,d) \right| \leq \Theta \right)}{\Pr \left( \left| \pi {-} \beta(\varphi,d) \right| \leq \Theta \right)} = \frac{{\rm P}_1(x)}{{\rm P}_2} ,
\end{align}
and ${\rm P}_1(x)$ is given as follows
\begin{align}
{\rm P}_1(x) & = \Pr \left( \frac{\cos^2 \left( \beta(\varphi,d) \right)}{\upsilon(d)} \leq x , \left| \pi {-} \beta(\varphi,d) \right| \leq \Theta \right) , \label{app:cdfuno_p1_1}\\
&= \int_{d_{\rm min}}^{d_{\rm max}} \Pr \Big( \cos \left( 2\beta(\varphi,r) \right) \leq 2x\upsilon(r)-1, \left| \pi {-} \beta(\varphi,r) \right| \leq \Theta \Big) {\rm d}r ,\label{app:cdfuno_p1_2}
\end{align}
where we drop the coefficient $1/\Delta d$ in \eqref{app:cdfuno_p1_2} as it would appear appear in ${\rm P}_2$, and will therefore be canceled eventually in \eqref{app:cdf_unordered_1}. Note that whenever $x \,{\geq}\, 1/\upsilon(r)$, \eqref{app:cdfuno_p1_2} reduces to
\begin{align} 
{\rm P}_1(x) \,{=} \int_{d_{\rm min}}^{d_{\rm max}} \!\!\! \Pr \left( \pi {-} \tan^{{-}1}(\ell/r) {-} \Theta \,{\leq}\, \varphi \,{\leq}\, \pi {-} \tan^{{-}1}(\ell/r) {+} \Theta \right) {\rm d}r \,{=} \int_{d_{\rm min}}^{d_{\rm max}} \!\!\! \Delta F_{\varphi} \left( r,\Theta \right) {\rm d}r. \label{app:cdfuno_p1_3} 
\end{align}

\begin{figure}[!t]
\centering
\includegraphics[width=0.75\textwidth]{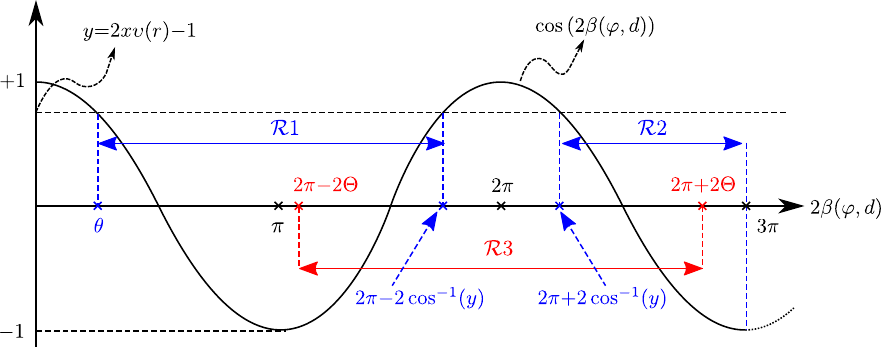}\vspace{-0.15in}
\caption{A sketch for the variation of $\cos \left( 2\beta(\varphi,r) \right)$.}
\label{fig:varphi}\vspace{-0.35in}
\end{figure}

On the other hand, whenever we have $0 \,{\leq}\, x \,{<}\, 1/\upsilon(r)$, we need to figure out the support of $\varphi$ satisfying both the conditions in the probability expression of \eqref{app:cdfuno_p1_2} simultaneously. In Fig.~\ref{fig:varphi}, we sketch the non-monotonic variation of $\cos \left( 2\beta(\varphi,r) \right)$. We observe that the first condition $\cos \left( 2\beta(\varphi,r) \right) \,{\leq}\, 2x\upsilon(r){-}1$ of \eqref{app:cdfuno_p1_2} is satisfied whenever $2\beta(\varphi,r)$ takes a value in one of the disconnected intervals $\mathcal{R}1$ and $\mathcal{R}2$. Moreover, in order to satisfy the second condition $\pi-\Theta \leq \beta(\varphi,d) \leq \pi+\Theta$ of \eqref{app:cdfuno_p1_2}, as well, $2\beta(\varphi,r)$ should be in the interval $\mathcal{R}3$. As a result, both these conditions are satisfied at the same time if $2\beta(\varphi,r) \,{\in}\, \mathcal{R}$ with $\mathcal{R} \,{=}\, \left( \mathcal{R}1 \cup \mathcal{R}2 \right) \cap \mathcal{R}3$. By Fig.~\ref{fig:varphi}, we have $\mathcal{R} \,{=}\, \varnothing$ whenever $1/2\cos^{{-}1}\!\left( 2x\upsilon(r){-}1\right) \,{>}\, \Theta$, or equivalently $x\,{\leq} \cos^2 \Theta {/}\upsilon(r)$, which corresponds to the receive direction being outside the FOV. As a result, \eqref{app:cdfuno_p1_2} becomes 
\begin{align} 
{\rm P}_1(x) & = \int_{d_{\rm min}}^{d_{\rm max}} \Big[ \Pr \left( \pi-\Theta \leq \beta(\varphi,r) \leq \pi- 1/2\cos^{{-}1}\!\left( 2x\upsilon(r){-}1\right)\right) \nonumber \\
& \qquad + \Pr \left( \pi+ 1/2\cos^{{-}1}\!\left( 2x\upsilon(r){-}1\right) \leq \beta(\varphi,r) \leq \pi + \Theta \right)  \big] {\rm d}r, \label{app:cdfuno_p1_4} \\
&= \int_{d_{\rm min}}^{d_{\rm max}} \left( \Delta F_{\varphi} \left( r,\Theta \right) - \Delta F_{\varphi} \left( r,1/2\cos^{{-}1}\!\left( 2x\upsilon(r){-}1\right) \right) \right) {\rm d}r,\label{app:cdfuno_p1_5}
\end{align}
if $\cos^2 \Theta {/}\upsilon(r) \,{<}\,x\,{<}\, 1/\upsilon(r)$, and ${\rm P}_1(x) \,{=}\, 0$ whenever $x\,{\leq}\, \cos^2 \Theta {/}\upsilon(r)$. Using \eqref{app:cdfuno_p1_3} and \eqref{app:cdfuno_p1_5}, ${\rm P}_1(x)$ can be represented for any $x\,{\geq}\,0$ as follows
\begin{align} 
{\rm P}_1(x) & = \int_{d_{\rm min}}^{d_{\rm max}} \left( \Delta F_{\varphi} \left( r,\Theta \right) - \Delta F_{\varphi} \left( r,\psi(x,r,\Theta) \right) \right) {\rm d}r , \label{app:cdfuno_p1_6}
\end{align}
where $\psi(x,y,z)$ is given in Theorem~\ref{the:cdf_unordered}. Realizing that ${\rm P}_2$ is exactly given by \eqref{app:cdfuno_p1_3}, employing \eqref{app:cdfuno_p1_3} and \eqref{app:cdfuno_p1_6} in \eqref{app:cdf_unordered_1} yields \eqref{eqn:cdf_unordered}. \hfill\IEEEQEDhere

\vspace{-0.2in}
\section{Proof of Theorem~\ref{the:cdf_twobit_ins}}\label{app:cdf_twobit_ins}
The desired CDF of the user $i\,{\in}\,\mathcal{S}_{\rm W, \varphi}$ (i.e., $\varphi$ is available for feedback computation) is
\begin{align} 
F_{h_i^2|h_i>0}(x) &= \Pr \left( h^2 \leq x \,\big| h > 0, d > d_{\rm th}, \left|\theta\right| > \theta_{\rm th}  \right) \label{app:cdf_2bit_ins_1}\\
&= \frac{\Pr \left( h^2 \leq x , d > d_{\rm th}, \theta_{\rm th} < \left|\theta\right| \leq \Theta \right)}{\Pr \left( d > d_{\rm th}, \theta_{\rm th} < \left|\theta\right| \leq \Theta \right)} = \frac{{\rm P}_1(x)}{{\rm P}_2} ,\label{app:cdf_2bit_ins_2}
\end{align}
where the incidence angle $\theta$ is given in \eqref{eqn:incidenceang}, and \eqref{app:cdf_2bit_ins_2} makes use of the fact that nonzero channel gain is satisfied whenever the receive direction is inside FOV, i.e., $\left|\theta\right| \leq \Theta$. Following the strategy of Appendix~\ref{app:cdf_fullcsi} starting with \eqref{app:cdfuno_p1_1}, ${\rm P}_1(x)$ is given as
\begin{align}\label{app:cdf_2bit_ins_3}
{\rm P}_1(x) \,{=} \int_{d_{\rm th}}^{d_{\rm max}} \Pr \left(\frac{\cos^2 \left( \beta(\varphi,r) \right)}{\upsilon(r)} \leq x, \theta_{\rm th} < \left|\pi - \beta(\varphi,r)\right| \leq \Theta \right) {\rm d}r, 
\end{align}
where the only difference with the probability expression is an additional lower limit $\theta_{\rm th}$ for the term $\left|\pi {-} \beta(\varphi,r)\right|$. The interval $\mathcal{R}_3$ is therefore split into two disconnected sub-intervals, i.e., $\left[2\pi{-}2\Theta,2\pi{-}2\theta_{\rm th}\right]$ and $\left[2\pi{+}2\theta_{\rm th},2\pi{+}2\Theta\right]$. As a result, whenever $x \,{\geq}\, 1/\upsilon(r)$, \eqref{app:cdf_2bit_ins_3} reduces to
\begin{align} \label{app:cdf_2bit_ins_4}
{\rm P}_1(x) \,{=} \int_{d_{\rm th}}^{d_{\rm max}} \!\!\! \Pr \left(\theta_{\rm th} < \left|\pi - \beta(\varphi,r)\right| \leq \Theta \right) {\rm d}r \,{=} \int_{d_{\rm th}}^{d_{\rm max}} \!\!\! \left( \Delta F_{\varphi} \left( r,\Theta \right) - \Delta F_{\varphi} \left( r, \theta_{\rm th} \right) \right) {\rm d}r .
\end{align}  
On the other hand, when $\cos^2 \Theta {/}\upsilon(r) \,{<}\,x\,{<}\, 1/\upsilon(r)$, \eqref{app:cdf_2bit_ins_3}  turns out to be
\begin{align} 
{\rm P}_1(x) &= \int_{d_{\rm th}}^{d_{\rm max}} \Big[ \Pr \left( \pi-\Theta \leq \beta(\varphi,r) \leq \pi- \max \left( 1/2\cos^{{-}1}\!\left( 2x\upsilon(r){-}1\right), \theta_{\rm th}\right) \right) . \nonumber \\
& \qquad + \Pr \left( \pi+ \max \left( 1/2\cos^{{-}1}\!\left( 2x\upsilon(r){-}1\right), \theta_{\rm th}\right) \leq \beta(\varphi,r) \leq \pi + \Theta \right) \Big] {\rm d}r , \label{app:cdf_2bit_ins_5}\\
&= \int_{d_{\rm th}}^{d_{\rm max}} \Big[ \Delta F_{\varphi} \left( r,\Theta \right) - \Delta F_{\varphi} \left( r, \max \left( 1/2\cos^{{-}1}\!\left( 2x\upsilon(r){-}1\right), \theta_{\rm th}\right) \right),\label{app:cdf_2bit_ins_6}
\end{align}
and ${\rm P}_1(x) \,{=}\, 0$ for $x \,{\leq}\,\cos^2 \Theta {/}\upsilon(r)$. Finally, ${\rm P}_1(x,r)$ can be represented for any $x\,{\geq}\,0$ as follows
\begin{align}
{\rm P}_1(x) &= \int_{d_{\rm th}}^{d_{\rm max}} \Pi \left[ \frac{\cos^2 \Theta}{x\,\upsilon(r)}\right] \Big( \Delta F_{\varphi} \left( r,\Theta \right) - \Delta F_{\varphi} \left( r,\omega \left(x,r,\theta_{\rm th}\right) \right) \Big) {\rm d}r , \label{app:cdf_2bit_ins_7}
\end{align}
where $\omega(x,y,z) \,{=}\, \max \left( 1/2\cos^{{-}1}\!\left( 2\min \left(x\upsilon(y),1\right){-}1\right),z \right)$ and the term involving $\Pi$ function excludes the condition $x \,{<}\,\cos^2 \Theta {/}\upsilon(r)$. Note that \eqref{app:cdf_2bit_ins_7} can be equivalently written as
\begin{align}
{\rm P}_1(x) &= \int_{d^*(x)}^{d_{\rm max}} \Big( \Delta F_{\varphi} \left( r,\Theta \right) - \Delta F_{\varphi} \left( r,\omega \left(x,r,\theta_{\rm th}\right) \right) \Big) {\rm d}r , \label{app:cdf_2bit_ins_8}
\end{align}
where $d^*(x)$ is given in Theorem~\ref{the:cdf_twobit_ins}. Realizing that ${\rm P}_2$ is exactly given by \eqref{app:cdf_2bit_ins_4}, incorporating \eqref{app:cdf_2bit_ins_4} and \eqref{app:cdf_2bit_ins_8} into \eqref{app:cdf_2bit_ins_2} yields \eqref{eqn:cdf_twobit_ins_i}.\hfill\IEEEQEDhere

Similarly, the desired CDF of the user $j\,{\in}\,\mathcal{S}_{\rm S, \varphi}$ is given as
\begin{align} 
F_{h_j^2|h_j>0}(x) &= \Pr \left( h^2 \leq x \,\big| h > 0, d \leq d_{\rm th}, \left|\theta\right| \leq \theta_{\rm th}  \right) = \frac{\Pr \left( h^2 \leq x , d \leq d_{\rm th}, \left|\theta\right| \leq \theta_{\rm th} \right)}{\Pr \left( d \leq d_{\rm th}, \left|\theta\right| \leq \theta_{\rm th} \right)} ,\label{eqn:cdf2bit_ins_j2}
\end{align}
where we assume that $\theta_{\rm th} \,{\leq}\, \Theta$. Note that \eqref{eqn:cdf2bit_ins_j2} has a similar expression of \eqref{app:cdf_unordered_1} except $\Theta$ is substituted by $\theta_{\rm th}$, and the upper limit of distance is now $d_{\rm th}$ instead of $d_{\max}$. As a result, replacing $\theta_{\rm th}$ and $d_{\rm th}$ with $\Theta$ and $d_{\max}$, respectively, in \eqref{eqn:cdf_unordered} obtains \eqref{eqn:cdf_twobit_ins_j}.\hfill\IEEEQEDhere    

\vspace{-0.2in}
\section{Proof of Theorem~\ref{the:cdf_twobit_mean}}\label{app:cdf_twobit_mean}
The desired CDF of the user $i\,{\in}\,\mathcal{S}_{\rm W, \overline{\varphi}}$ (i.e., $\overline{\varphi}$ is available for feedback computation) is
\begin{align} 
F_{h_i^2|\overline{h}_i>0}(x) 
&= \frac{\Pr \left( h^2 \leq x , d > d_{\rm th}, \theta_{\rm th} < \left|\overline{\theta}\right| \leq \Theta \right)}{\Pr \left( d > d_{\rm th}, \theta_{\rm th} < \left|\overline{\theta}\right| \leq \Theta \right)} = \frac{{\rm P}_1(x)}{{\rm P}_2} ,\label{app:cdf_2bit_mea_2}
\end{align}
where $\overline{\theta} \,{=}\, \pi {-} \beta(\overline{\varphi},d)$. Considering \eqref{app:square_channel} and \eqref{app:cdf_2bit_ins_3}, ${\rm P}_1(x)$ is given as
\begin{align}
{\rm P}_1(x) &= \int_{d_{\rm th}}^{d_{\rm max}} \Pr \left( \frac{\cos^2 \left( \beta(\varphi,r) \right)}{\upsilon(r)} \, \Pi\left[ \left| \pi {-} \beta(\varphi,r) \right| /\Theta \right] \leq x, \theta_{\rm th} \leq \left|\pi {-} \beta(\overline{\varphi},r)\right| \leq \Theta \right) {\rm d}r, \label{app:cdf_2bit_mea_3}\\
&= \int_{d_{\rm th}}^{d_{\rm max}} \Big[ \Pr \left( \cos^2 \left( \beta(\varphi,r) \right) \leq x\,\upsilon(r) , \left| \pi {-} \beta(\varphi,d) \right| \leq \Theta, \theta_{\rm th} \leq \left|\pi {-} \beta(\overline{\varphi},r)\right| \leq \Theta \right) {\rm d}r \nonumber \\
&\hspace{0.85in} + \Pr \left( \left| \pi {-} \beta(\varphi,r) \right| > \Theta, \theta_{\rm th} \leq \left|\pi {-} \beta(\overline{\varphi},r)\right| \leq \Theta \right) \Big] {\rm d}r . \label{app:cdf_2bit_mea_4}
\end{align}
Defining $\Delta F_{\overline{\varphi}} \left( x,y \right) \,{=}\, F_{\overline{\varphi}}\left( \pi {-} \tan^{{-}1}(\ell/x) {+} y \right) \,{-}\, F_{\overline{\varphi}} \left( \pi {-} \tan^{{-}1}(\ell/x) {-} y  \right)$ with $F_{\overline{\varphi}}$ being the CDF of $\overline{\varphi}$, which is uniform with $\mathcal{U}\,[\overline{\varphi}_\text{min}{,}\,\overline{\varphi}_\text{max}]$, \eqref{app:cdf_2bit_mea_4} reduces to 
\begin{align}
{\rm P}_1(x) = \int_{d_{\rm th}}^{d_{\rm max}} \Pr \left( \theta_{\rm th} \leq \left|\pi {-} \beta(\overline{\varphi},r)\right| \leq \Theta \right) {\rm d}r = \int_{d_{\rm th}}^{d_{\rm max}} \left( \Delta F_{\overline{\varphi}} \left( r,\Theta \right) - \Delta F_{\overline{\varphi}} \left( r, \theta_{\rm th} \right) \right) {\rm d}r , \label{app:cdf_2bit_mea_5}
\end{align}
for $x \,{\geq}\, 1/\upsilon(r)$. In addition, whenever we have $x \,{<}\, 1/\upsilon(r)$, \eqref{app:cdf_2bit_mea_4} becomes 
\begin{align}
{\rm P}_1(x) &= \frac{1}{\Delta \overline{\varphi}} \int_{d_{\rm th}}^{d_{\rm max}} \int_{\mathcal{S}_{\overline{\varphi}}(r)} \Big[ \Pr \left( \cos^2 \left( \beta(\varphi,r) \right) \leq x\,\upsilon(r) , \left| \pi {-} \beta(\varphi,r) \right| \leq \Theta \,\big|\, \overline{\varphi} \right) {\rm d}r \nonumber \\
&\hspace{1.6in} + \Pr \left( \left| \pi {-} \beta(\varphi,r) \right| > \Theta \,\big|\, \overline{\varphi}  \right) \Big] {\rm d}\overline{\varphi} \, {\rm d}r , \label{app:cdf_2bit_mea_6}
\end{align}
where $\mathcal{S}_{\overline{\varphi}}(r)$ is the support of $\overline{\varphi}$ satisfying $\theta_{\rm th} \,{<}\, \left|\pi {-} \beta(\overline{\varphi},r)\right| \,{\leq}\, \Theta$ and $\overline{\varphi}_\text{min} \,{\leq}\,\overline{\varphi} \,{\leq}\, \overline{\varphi}_\text{max}$ jointly, and is given in Theorem~\ref{the:cdf_twobit_mean}. Note that the first probability in \eqref{app:cdf_2bit_mea_6} is very similar to the one in \eqref{app:cdfuno_p1_1} except the distribution of $\varphi$, which is conditional in \eqref{app:cdf_2bit_mea_6} with a given $\overline{\varphi}$, and is unconditional in \eqref{app:cdfuno_p1_1}. Note also that the second probability in \eqref{app:cdf_2bit_mea_6} is actually the complement of the one in \eqref{app:cdfuno_p1_3}. As a result, using \eqref{app:cdfuno_p1_3} and \eqref{app:cdfuno_p1_6} (i.e., the final form of \eqref{app:cdfuno_p1_1}), \eqref{app:cdf_2bit_mea_6} can be represented as
\begin{align}
{\rm P}_1(x) \,{=}\, \frac{1}{\Delta \overline{\varphi}} \int_{d_{\rm th}}^{d_{\rm max}} \!\!\!\! \int_{\mathcal{S}_{\overline{\varphi}}(r)} \Big[ \underbrace{\Delta F_{\varphi|\overline{\varphi}} \left( r,\Theta \right) {-} \Delta F_{\varphi|\overline{\varphi}} \left( r,\Psi(x,r,\Theta) \right)}_{1\textrm{st probability in \eqref{app:cdf_2bit_mea_6}}}  {+} \underbrace{1 {-} \Delta F_{\varphi|\overline{\varphi}} \left( r,\Theta \right)}_{2\textrm{nd probability in \eqref{app:cdf_2bit_mea_6}}} \Big] {\rm d}\overline{\varphi} \, {\rm d}r , \label{app:cdf_2bit_mea_7}
\end{align}
where we use $\Psi(x,y,z)$ instead of $\psi(x,y,z)$ (in Theorem~\ref{the:cdf_unordered}) since this particular case considers only $x \,{<}\, 1/\upsilon(r)$ (i.e., not any $x \,{\geq}\, 0$ as for $\psi(x,y,z)$). Considering \eqref{app:cdf_2bit_mea_5} and \eqref{app:cdf_2bit_mea_7}, ${\rm P}_1(x)$ becomes
\begin{align}
\!\!\! {\rm P}_1(x) \,{=} \!\!\int_{d^*(x)}^{d_{\rm max}} \!\!\! \left( \Delta F_{\overline{\varphi}} \left( r,\Theta \right) {-} \Delta F_{\overline{\varphi}} \left( r, \theta_{\rm th} \right) \right) {\rm d}r + \frac{1}{\Delta \overline{\varphi}} \!\! \int_{d_{\rm th}}^{d^*(x)} \!\!\!\!\! \int_{\mathcal{S}_{\overline{\varphi}}(r)} \!\!\! \!\!\!\!\!\! \left( 1{-} \Delta F_{\varphi|\overline{\varphi}} \left( r,\Psi(x,r,\Theta) \right) \right)  {\rm d}\overline{\varphi} \, {\rm d}r,  \label{app:cdf_2bit_mea_8} 
\end{align}
for any $x\,{\geq}\,0$, where we adjust the integral limits by including $d^*(x)$ (defined in Theorem~\ref{the:cdf_twobit_mean}) to capture the contributions of the conditions $x \,{<}\, 1/\upsilon(r)$ and $x \,{\geq}\, 1/\upsilon(r)$, which is a similar strategy of \eqref{app:cdf_2bit_ins_8}. Realizing that ${\rm P}_2$ is exactly given by \eqref{app:cdf_2bit_mea_5}, the desired CDF in \eqref{eqn:cdf_twobit_mean_i} is readily obtained by employing \eqref{app:cdf_2bit_mea_5} and \eqref{app:cdf_2bit_mea_8} in \eqref{app:cdf_2bit_mea_2} with the definition
\begin{align}\label{app:auxiliary_a}
A(x) = \int_{x}^{d_{\rm max}} \Big( \Delta F_{\overline{\varphi}} \left( r,\Theta \right) - \Delta F_{\overline{\varphi}} \left( r,\theta_{\rm th} \right) \Big) {\rm d}r.
\end{align}
Moreover, defining $I(x,y,z)$ to be
\begin{align}\label{app:integral_defn}
I(x,y,z) \,{=}\, \int_{y}^{z} F_{\overline{\varphi}} \left( \pi - \tan^{{-}1}(\ell/r) + x \right) {\rm d}r ,
\end{align}
and employing \eqref{app:integral_defn} in \eqref{app:auxiliary_a}, we readily obtain \eqref{eqn:auxiliary_a}. Using the definition 
\begin{align}\label{app:cdf_meanvarphi}
F_{\overline{\varphi}} \left( \pi - \tan^{{-}1}(\ell/r) + x \right) = \begin{cases}
1 &  \textrm{ if } \tan^{{-}1}(\ell/r) \leq \pi {+} x {-} \overline{\varphi}_{\max} ,\\
\displaystyle g^{'}(x,r) &  \textrm{ if } \pi {+} x {-} \overline{\varphi}_{\max} < \tan^{{-}1}(\ell/r) \leq \pi {+} x {-} \overline{\varphi}_{\min} ,\\
0 & \textrm{ if } \tan^{{-}1}(\ell/r) > \pi {+} x {-} \overline{\varphi}_{\min},
\end{cases}
\end{align}
with $g^{'}(x,r) \,{=}\, \left( \pi \,{+}\, x \,{-}\, \overline{\varphi}_{\min} {-} \tan^{{-}1}(\ell/r) \right)/\Delta\overline{\varphi} $, and the geometrical relation $0\,{\leq}\,\tan^{{-}1}(\ell/d)\,{<}\,\pi/2$, we will derive the closed form expression of $I(x,y,z)$ in the following. Before that, we define 
\begin{align}\label{app:antiderivative}
\int_{a}^{b} g^{'}(x,r) \, {\rm d}r = g(x,r)|_{r=b} - g(x,r)|_{r=a} = \Delta g(x,a,b),
\end{align}
where $g(x,r)$ is the antiderivative of $g^{'}(x,r)$, which is given in \eqref{eqn:exact_integral} and computed using \cite[($2.854$)]{Gradshteyn}. In the derivation, we use \eqref{app:antiderivative} together with the definitions $r_{\min}(x)$, $r_{\max}(x)$, and $\alpha(x,r)$ of Theorem~\ref{the:cdf_twobit_mean}, and $u(x,y,z)$ of Theorem~\ref{the:cdf_twobit_ins}. In addition, all the conditions in the following case definitions capture the lower and upper limits of the inequalities in \eqref{app:cdf_meanvarphi} and the geometry-based boundaries of $\tan^{{-}1}(\ell/d)$ (i.e., $[0,\pi/2]$) jointly.

\paragraph{Case 1} $\pi \,{+}\, x \,{-}\, \overline{\varphi}_{\max} \,{<}\, 0$, $\pi \,{+}\, x \,{-}\, \overline{\varphi}_{\min} \,{<}\, 0$ $ \longrightarrow I(x,y,z) = 0.$ 

\paragraph{Case 2} $\pi \,{+}\, x \,{-}\, \overline{\varphi}_{\max} \,{<}\, 0$, $0 \,{\leq}\, \pi \,{+}\, x \,{-}\, \overline{\varphi}_{\min} \,{<}\, \pi/2$ 
\begin{align}
I(x,y,z) = \int_{u(r_{\min}(x),y,z)}^{z} g^{'}(x,r) \, {\rm d}r = \Delta g(x,u(r_{\min}(x),y,z),z) .
\end{align}

\paragraph{Case 3} $\pi \,{+}\, x \,{-}\, \overline{\varphi}_{\max} \,{<}\, 0$, $\pi \,{+}\, x \,{-}\, \overline{\varphi}_{\min} \,{\geq}\, \pi/2$ 
\begin{align}
I(x,y,z) = \int_{y}^{z} g^{'}(x,r) \, {\rm d}r = \Delta g(x,y,z) . 
\end{align}

\paragraph{Case 4} $0 \,{\leq}\, \pi \,{+}\, x \,{-}\, \overline{\varphi}_{\max} \,{<}\, \pi/2$, $0 \,{\leq}\, \pi \,{+}\, x \,{-}\, \overline{\varphi}_{\min} \,{<}\, \pi/2$ 
\begin{align}
I(x,y,z) &=  \int_{u(r_{\max}(x),y,z)}^{z} \, {\rm d}r + \int_{u(r_{\min}(x),y,z)}^{u(r_{\max}(x),y,z)} g^{'}(x,r) \, {\rm d}r , \\
&= z - u(r_{\max}(x),y,z) + \Delta g(x,u(r_{\min}(x),y,z),u(r_{\max}(x),y,z)) .
\end{align}

\paragraph{Case 5} $0 \,{\leq}\, \pi \,{+}\, x \,{-}\, \overline{\varphi}_{\max} \,{<}\, \pi/2$, $\pi \,{+}\, x \,{-}\, \overline{\varphi}_{\min} \,{\geq}\, \pi/2$ 
\begin{align}
I(x,y,z) &= \int_{u(r_{\max}(x),y,z)}^{z} \, {\rm d}r + \int_{y}^{u(r_{\max}(x),y,z)} g^{'}(x,r) \, {\rm d}r , \\
&= z - u(r_{\max}(x),y,z) + \Delta g(x,y,u(r_{\max}(x),y,z)) .
\end{align}

\paragraph{Case 6} $\pi \,{+}\, x \,{-}\, \overline{\varphi}_{\max} \,{\geq}\, \pi/2$, $\pi \,{+}\, x \,{-}\, \overline{\varphi}_{\min} \,{\geq}\, \pi/2$ $\longrightarrow I(x,y,z) = z - y.$

After employing the expressions of $I(x,y,z)$ in Case $1{-}6$, we readily obtain \eqref{eqn:integral}.\hfill\IEEEQEDhere

Similarly, the desired CDF of the user $j\,{\in}\,\mathcal{S}_{\rm W, \overline{\varphi}}$ (i.e., $\overline{\varphi}$ is available for feedback) is
\begin{align} 
F_{h_j^2|\overline{h}_j>0}(x) 
&= \frac{\Pr \left( h^2 \leq x , d \leq d_{\rm th}, \left|\overline{\theta}\right| \leq \theta_{\rm th} \right)}{\Pr \left( d \leq d_{\rm th}, \left|\overline{\theta}\right| \leq \theta_{\rm th} \right)} = \frac{{\rm P}_1(x)}{{\rm P}_2},\label{app:cdf_2bit_mea_j2}
\end{align}
where ${\rm P}_1(x)$ is given similar to \eqref{app:cdf_2bit_mea_4} as follows
\begin{align}
{\rm P}_1(x) &= \int_{d_{\min}}^{d_{\rm th}} \Big[ \Pr \left( \cos^2 \left( \beta(\varphi,r) \right) \leq x\,\upsilon(r) , \left| \pi {-} \beta(\varphi,r) \right| \leq \Theta, \left|\pi {-} \beta(\overline{\varphi},r)\right| \leq \theta_{\rm th} \right) {\rm d}r \nonumber \\
&\hspace{0.85in} + \Pr \left( \left| \pi {-} \beta(\varphi,r) \right| > \Theta, \left|\pi {-} \beta(\overline{\varphi},r)\right| \leq \theta_{\rm th} \right) \Big] {\rm d}r . \label{app:cdf_2bit_mea_j3}
\end{align}
With $B(x)$ given in Theorem~\ref{the:cdf_twobit_mean}, \eqref{app:cdf_2bit_mea_j3} for $x \,{\geq}\, 1/\upsilon(r)$ becomes
\begin{align}
{\rm P}_1(x) = \int_{d_{\min}}^{d_{\rm th}} \Pr \left( \left|\pi {-} \beta(\overline{\varphi},r)\right| \leq \theta_{\rm th}  \right) {\rm d}r = \int_{d_{\min}}^{d_{\rm th}} \Delta F_{\overline{\varphi}} \left( r,\theta_{\rm th} \right) {\rm d}r = B(d_{\min}), \label{app:cdf_2bit_mea_j4}
\end{align}
Following the strategy of \eqref{app:cdf_2bit_mea_7}, \eqref{app:cdf_2bit_mea_j3} is computed for $x \,{<}\, 1/\upsilon(r)$ to be
\begin{align}
{\rm P}_1(x) = \frac{1}{\Delta \overline{\varphi}} \int_{d_{\min}}^{d_{\rm th}} \int_{\overline{\varphi}_2(r)}^{\overline{\varphi}_3(r)} \left( 1{-} \Delta F_{\varphi|\overline{\varphi}} \left( r,\Psi(x,r,\Theta) \right) \right)  {\rm d}\overline{\varphi} \, {\rm d}r.\label{app:cdf_2bit_mea_j5}
\end{align}
Using $d^*(x)$ in the integration limits of \eqref{app:cdf_2bit_mea_j4} and \eqref{app:cdf_2bit_mea_j5} (as in \eqref{app:cdf_2bit_mea_8}), ${\rm P}_1(x)$ for any $x\,{\geq}\,0$ becomes
\begin{align}
{\rm P}_1(x) = B(d^*(x)) + \frac{1}{\Delta \overline{\varphi}} \int_{d_{\min}}^{d^*(x)} \int_{\overline{\varphi}_2(r)}^{\overline{\varphi}_3(r)} \left( 1{-} \Delta F_{\varphi|\overline{\varphi}} \left( r,\Psi(x,r,\Theta) \right) \right)  {\rm d}\overline{\varphi} \, {\rm d}r.\label{app:cdf_2bit_mea_j6}
\end{align}
Realizing that ${\rm P}_2$ is given by \eqref{app:cdf_2bit_mea_j4}, combining \eqref{app:cdf_2bit_mea_j4} and \eqref{app:cdf_2bit_mea_j6} in \eqref{app:cdf_2bit_mea_j2} yields \eqref{eqn:cdf_twobit_mean_j}.\hfill\IEEEQEDhere

\vspace{-0.2in}
\bibliographystyle{IEEEtran}
\bibliography{IEEEabrv,papers}

\begin{thebibliography}{10}
\providecommand{\url}[1]{#1}
\csname url@samestyle\endcsname
\providecommand{\newblock}{\relax}
\providecommand{\bibinfo}[2]{#2}
\providecommand{\BIBentrySTDinterwordspacing}{\spaceskip=0pt\relax}
\providecommand{\BIBentryALTinterwordstretchfactor}{4}
\providecommand{\BIBentryALTinterwordspacing}{\spaceskip=\fontdimen2\font plus
\BIBentryALTinterwordstretchfactor\fontdimen3\font minus
  \fontdimen4\font\relax}
\providecommand{\BIBforeignlanguage}[2]{{%
\expandafter\ifx\csname l@#1\endcsname\relax
\typeout{** WARNING: IEEEtran.bst: No hyphenation pattern has been}%
\typeout{** loaded for the language `#1'. Using the pattern for}%
\typeout{** the default language instead.}%
\else
\language=\csname l@#1\endcsname
\fi
#2}}
\providecommand{\BIBdecl}{\relax}
\BIBdecl

\bibitem{Richardson13VLC}
A.~Jovicic, J.~Li, and T.~Richardson, ``Visible light communication:
  {O}pportunities, challenges and the path to market,'' \emph{IEEE Commun.
  Mag.}, vol.~51, no.~12, pp. 26--32, Dec. 2013.

\bibitem{Yapici2017MulEleVLC}
Y.~S. Ero\u{g}lu, I.~G\"{u}ven\c{c}, A.~\c{S}ahin, Y.~Yap{\i}c{\i}
  \emph{et~al.}, ``Multi-element {VLC} networks: {LED} assignment, power
  control, and optimum combining,'' \emph{{IEEE} J. Sel. Areas Commun.},
  vol.~36, no.~1, pp. 121--135, Jan. 2018.

\bibitem{Eroglu2018SofDef}
S.~I. Mushfique, P.~Palathingal, Y.~S. Eroglu, M.~Yuksel \emph{et~al.}, ``A
  software-defined multi-element {VLC} architecture,'' \emph{IEEE Commun.
  Mag.}, vol.~56, no.~2, pp. 196--203, Feb. 2018.

\bibitem{Haas14VLCBeyond}
H.~Burchardt, N.~Serafimovski, D.~Tsonev, S.~Videv \emph{et~al.}, ``{VLC}:
  {B}eyond point-to-point communication,'' \emph{IEEE Commun. Mag.}, vol.~52,
  no.~7, pp. 98--105, Jul. 2014.

\bibitem{Haas2014GbpsVLC}
D.~Tsonev, H.~Chun, S.~Rajbhandari, J.~J.~D. McKendry \emph{et~al.}, ``{A
  3-Gb/s Single-LED OFDM-Based Wireless VLC Link Using a Gallium Nitride
  $\mu{\rm LED}$},'' \emph{IEEE Photon. Technol. Lett.}, vol.~26, no.~7, pp.
  637--640, Apr. 2014.

\bibitem{Ding2017AppNoma}
Z.~Ding, Y.~Liu, J.~Choi, Q.~Sun \emph{et~al.}, ``Application of non-orthogonal
  multiple access in {LTE and 5G} networks,'' \emph{IEEE Commun. Mag.},
  vol.~55, no.~2, pp. 185--191, Feb. 2017.

\bibitem{Poor2017NomaMul}
W.~Shin, M.~Vaezi, B.~Lee, D.~J. Love \emph{et~al.}, ``Non-orthogonal multiple
  access in multi-cell networks: Theory, performance, and practical
  challenges,'' \emph{IEEE Commun. Mag.}, vol.~55, no.~10, pp. 176--183, Oct.
  2017.

\bibitem{Dobre2017PowDomNoma}
S.~M.~R. Islam, N.~Avazov, O.~A. Dobre, and K.~s.~Kwak, ``Power-domain
  non-orthogonal multiple access {(NOMA)} in {5G} systems: {Potentials} and
  challenges,'' \emph{IEEE Commun. Surveys Tuts.}, vol.~19, no.~2, pp.
  721--742, 2nd Quarter 2017.

\bibitem{Uysal2015Noma}
R.~C. Kizilirmak, C.~R. Rowell, and M.~Uysal, ``Non-orthogonal multiple access
  {(NOMA)} for indoor visible light communications,'' in \emph{Proc. Int.
  Workshop on Optical Wireless Commun. (IWOW'2015)}, Sep. 2015, pp. 98--101.

\bibitem{Haas2016PerEvaNoma}
L.~Yin, W.~O. Popoola, X.~Wu, and H.~Haas, ``Performance evaluation of
  non-orthogonal multiple access in visible light communication,'' \emph{IEEE
  Trans. Commun.}, vol.~64, no.~12, pp. 5162--5175, Dec. 2016.

\bibitem{Karagiannidis2016Noma}
H.~Marshoud, V.~M. Kapinas, G.~K. Karagiannidis, and S.~Muhaidat,
  ``Non-orthogonal multiple access for visible light communications,''
  \emph{IEEE Photon. Technol. Lett.}, vol.~28, no.~1, pp. 51--54, Jan. 2016.

\bibitem{Du2017OnPeMimo}
C.~Chen, W.~D. Zhong, H.~Yang, and P.~Du, ``On the performance of {MIMO-NOMA}
  based visible light communication systems,'' \emph{IEEE Photon. Technol.
  Lett.}, vol.~PP, no.~99, pp. 1--1, Dec. 2017.

\bibitem{Karagiannidis2017OnPerNoma}
H.~Marshoud, P.~C. Sofotasios, S.~Muhaidat, G.~K. Karagiannidis \emph{et~al.},
  ``On the performance of visible light communication systems with
  non-orthogonal multiple access,'' \emph{IEEE Trans. Wireless Commun.},
  vol.~16, no.~10, pp. 6350--6364, Oct. 2017.

\bibitem{Li2017FaiNoma}
Z.~Yang, W.~Xu, and Y.~Li, ``Fair non-orthogonal multiple access for visible
  light communication downlinks,'' \emph{IEEE Wireless Commun. Lett.}, vol.~6,
  no.~1, pp. 66--69, Feb. 2017.

\bibitem{Xu2017UseGro}
X.~Zhang, Q.~Gao, C.~Gong, and Z.~Xu, ``User grouping and power allocation for
  {NOMA} visible light communication multi-cell networks,'' \emph{IEEE Commun.
  Lett.}, vol.~21, no.~4, pp. 777--780, Apr. 2017.

\bibitem{Cha2017JoiDet}
X.~Guan, Q.~Yang, and C.~K. Chan, ``Joint detection of visible light
  communication signals under non-orthogonal multiple access,'' \emph{IEEE
  Photon. Technol. Lett.}, vol.~29, no.~4, pp. 377--380, Feb. 2017.

\bibitem{Haas2016AccPoiSel}
M.~D. Soltani, X.~Wu, M.~Safari, and H.~Haas, ``Access point selection in
  {Li-Fi} cellular networks with arbitrary receiver orientation,'' in
  \emph{Proc. {IEEE} Int. Symp. Pers. Indoor Mobile Radio Commun. (PIMRC)},
  Sep. 2016, pp. 1--6.

\bibitem{Haas2017HanModInd}
M.~D. Soltani, H.~Kazemi, M.~Safari, and H.~Haas, ``Handover modeling for
  indoor {Li-Fi} cellular networks: {The} effects of receiver mobility and
  rotation,'' in \emph{Proc. {IEEE} Wireless Commun. Netw. Conf. (WCNC)}, Mar.
  2017, pp. 1--6.

\bibitem{Yapici2017VerOriC}
Y.~S. Ero\u{g}lu, Y.~Yap{\i}c{\i}, and I.~G\"{u}ven\c{c}, ``Effect of random
  vertical orientation for mobile users in visible light communications,'' in
  \emph{Proc. Asilomar Conf. Signals, Syst., and Comput.}, Pacific Grove,
  California, Oct. 2017.

\bibitem{Yapici2017RanVerOr}
------, ``Impact of random receiver orientation on visible light communications
  channel,'' \emph{{IEEE} Trans. Commun.}, vol.~67, no.~2, pp. 1313--1325, Feb.
  2019.

\bibitem{Chen2017ImpBer}
J.~Y. Wang, J.~B. Wang, B.~Zhu, M.~Lin \emph{et~al.}, ``Improvement of {BER}
  performance by tilting receiver plane for indoor visible light communications
  with input-dependent noise,'' in \emph{Proc. {IEEE} Int. Conf. Commun.
  (ICC)}, May 2017, pp. 1--6.

\bibitem{Wang2017ImpRec}
J.~Y. Wang, Q.~L. Li, J.~X. Zhu, and Y.~Wang, ``Impact of receiver's tilted
  angle on channel capacity in {VLCs},'' \emph{IET Electron. Lett.}, vol.~53,
  no.~6, pp. 421--423, Mar. 2017.

\bibitem{Haas2018ImpTerOri}
A.~A. Purwita, M.~D. Soltani, M.~Safari, and H.~Haas, ``Impact of terminal
  orientation on performance in {LiFi} systems,'' in \emph{Proc. {IEEE}
  Wireless Commun. Netw. Conf. (WCNC)}, Apr. 2018, pp. 1--6.

\bibitem{Haas2018ModRanOri}
M.~D. {Soltani}, A.~A. {Purwita}, Z.~{Zeng}, H.~{Haas} \emph{et~al.},
  ``Modeling the random orientation of mobile devices: {Measurement}, analysis
  and {LiFi} use case,'' \emph{IEEE Trans. Commun. \normalfont{Early Access}},
  2018.

\bibitem{Ding2017RanBea}
Z.~Ding, P.~Fan, and H.~V. Poor, ``Random beamforming in millimeter-wave {NOMA}
  networks,'' \emph{IEEE Access}, vol.~5, pp. 7667--7681, Feb. 2017.

\bibitem{Haas2018PhySec}
L.~Yin and H.~Haas, ``Physical-layer security in multiuser visible light
  communication networks,'' \emph{IEEE J. Sel. Areas Commun.}, vol.~36, no.~1,
  pp. 162--174, Jan. 2018.

\bibitem{Moser2009OnCap}
A.~{Lapidoth}, S.~M. {Moser}, and M.~A. {Wigger}, ``On the capacity of
  free-space optical intensity channels,'' \emph{IEEE Trans. Inf. Theory},
  vol.~55, no.~10, pp. 4449--4461, Oct. 2009.

\bibitem{Poor2017RanBea}
Z.~Ding, P.~Fan, and H.~V. Poor, ``Random beamforming in millimeter-wave {NOMA}
  networks,'' \emph{IEEE Access}, vol.~5, no.~99, pp. 7667--7681, Feb. 2017.

\bibitem{Nagaraja2005OrdSta}
H.~A. David and H.~N. Nagaraja, \emph{Order Statistics}, 3rd~ed., ser. Wiley
  Series in Probability and Statistics.\hskip 1em plus 0.5em minus 0.4em\relax
  NJ: John Wiley \& Sons, Inc., Jan. 2005.

\bibitem{Nagaraja2009DisCon}
Q.~He and H.~Nagaraja, ``Distribution of concomitants of order statistics and
  their order statistics,'' \emph{J. Statist. Plann. Inference}, vol. 139,
  no.~8, pp. 2643--2655, Aug. 2009.

\bibitem{Yapici2018AngFee}
N.~Rupasinghe, Y.~Yap{\i}c{\i}, I.~G\"{u}ven\c{c}, M.~Ghosh \emph{et~al.},
  ``Angular feedback for {mmWave NOMA} drone networks,'' \emph{{\normalfont
  under review in} {IEEE} J. Sel. Topics Signal Process.}, Aug. 2018.

\bibitem{Yapici2018NOMAmmWDro_J}
N.~Rupasinghe, Y.~Yap{\i}c{\i}, I.~G\"{u}ven\c{c}, and Y.~Kakishima,
  ``Non-orthogonal multiple access for {mmWave} drone networks with limited
  feedback,'' \emph{{IEEE} Trans. Commun.}, vol.~67, no.~1, pp. 762--777, Jan.
  2019.

\bibitem{Lau2018SimPos}
B.~{Zhou}, V.~{Lau}, Q.~{Chen}, and Y.~{Cao}, ``Simultaneous positioning and
  orientating for visible light communications: {Algorithm} design and
  performance analysis,'' \emph{IEEE Trans. Vehic. Technol.}, vol.~67, no.~12,
  pp. 11\,790--11\,804, Dec. 2018.

\bibitem{Gradshteyn}
I.~S. Gradshteyn and I.~M. Ryzhik, \emph{Table of Integrals, Series, and
  Products}, 8th~ed., D.~Zwillinger and V.~Moll, Eds.\hskip 1em plus 0.5em
  minus 0.4em\relax Academic Press, 2014.

\end{thebibliography}

\end{document}